\documentclass[journal]{IEEEtran}
   \usepackage{graphicx}
\usepackage{multicol}
\usepackage[linesnumbered,ruled]{algorithm2e}
\usepackage[noend]{algpseudocode}
\usepackage{booktabs}
\usepackage{amssymb}
\usepackage{amsthm}
\usepackage{blkarray}
\usepackage{multirow} 
\usepackage[cmex10]{amsmath}
\usepackage{subcaption}
\usepackage{lipsum}
\usepackage{xcolor}
\usepackage{pdfpages}
\usepackage[noadjust]{cite}
  
\usepackage[left=0.64in, right=0.64in, top=0.75in, bottom=1.0in]{geometry}  
\usepackage{relsize}

\usepackage{graphicx} 
\usepackage{amsmath,bm} 
\usepackage{esvect}     
\usepackage{caption}
\usepackage{subcaption}

\usepackage{array}
\newcolumntype{M}[1]{>{\centering\arraybackslash}m{#1}}

\makeatletter
\let\old@ps@headings\ps@headings
\let\old@ps@IEEEtitlepagestyle\ps@IEEEtitlepagestyle
\def\confheader#1{%
	\def\ps@headings{%
		\old@ps@headings%
		\def\@oddhead{\strut\hfill#1\hfill\strut}%
		\def\@evenhead{\strut\hfill#1\hfill\strut}%
	}%
	\def\ps@IEEEtitlepagestyle{%
		\old@ps@IEEEtitlepagestyle%
		\def\@oddhead{\strut\hfill#1\hfill\strut}%
		\def\@evenhead{\strut\hfill#1\hfill\strut}%
	}%
	\ps@headings%
}
\def\endthebibliography{%
	\def\@noitemerr{\@latex@warning{Empty `thebibliography' environment}}%
	\endlist
}
\makeatother

\begin{document}


\title{An Artificial Intelligence Enabled Signature Estimation of Dual Wideband Systems in Ultra-Low Signal-to-Noise Ratio}

\author{Chandrashekhar Rai and Debarati Sen,
\thanks{Both the authors are with G.S.Sanyal School of Telecommunications, Indian Institute of Technology (IIT) Kharagpur, India. Email: \emph{cs.rai93@iitkgp.ac.in, debarati@gssst.iitkgp.ac.in}.}
\thanks{The conference precursor to this work has been published in the Vehicular Technology Conference (VTC) 2022.}
}

\markboth{IEEE Transactions on Cognitive Communications and Networking ,~Vol.~0, No.~0, 
April~2025}%
{Rai \MakeLowercase{\textit{et al.}}: A Sample Article Using IEEEtran.cls for IEEE Journals}

\maketitle

\begin{abstract}
Millimeter-wave (mmWave) massive Multiple Input Multiple Output (MIMO) systems encounter both spatial wideband spreading and temporal wideband effects in the communication channels of individual users. Accurate estimation of a user’s channel signature—specifically, the direction of arrival and time of arrival—is crucial for designing efficient beamforming transceivers, especially under noisy observations. In this work, we propose an Artificial Intelligence (AI)-enabled framework for estimating the channel signature of a user's location in mmWave massive MIMO systems. Our approach explicitly accounts for spatial wideband spreading, finite basis leakage effects, and significant unknown receiver noise. We demonstrate the effectiveness of a denoising convolutional neural network with residual learning for recovering channel responses, even when channel gains are of extremely low amplitude and embedded in ultra-high receiver noise environments. Notably, our method successfully recovers spatio-temporal diversity branches at signal-to-noise ratios as low as -20 dB. Furthermore, we introduce a local gravitation-based clustering algorithm to infer the number of physical propagation paths (unknown a priori) and to identify their respective support in the delay-angle domain of the denoised response. To complement our approach, we design tailored metrics for evaluating denoising and clustering performance within the context of wireless communications. We validate our framework through system-level simulations using Orthogonal Frequency Division Multiplexing (OFDM) with a Quadrature Phase Shift Keying (QPSK) modulation scheme over mmWave fading channels, highlighting the necessity and robustness of the proposed methods in ultra-low SNR scenarios.
\end{abstract}


\IEEEpeerreviewmaketitle
\section{Introduction} 
\label{section_Introduction}

\IEEEPARstart{M}{illimeter-wave} (mmWave) operations, combined with massive Multiple Input Multiple Output (MIMO) architectures \cite{lu2014overview, busari2017millimeter, hemadeh2017millimeter}, offer significant advantages such as enhanced spectral efficiency, high data rates, and ultra-high spatio-temporal resolution. However, these benefits come at the cost of increased complexity in both signal processing and system-level optimization, particularly due to wideband effects. In recent years, several efforts have been made to fully exploit the capabilities of mmWave massive MIMO systems under these wideband conditions \cite{brady2015wideband, xie2016unified, wang2018spatial}.

In such systems, signals from multiple reflecting or scattering objects arrive at the antenna array with different delays, influenced by the inter-element spacing and angle of arrival (AoA). These delays, along with the array geometry, define the spatial signature of the wireless channel. In communications, these spatial signatures enable efficient channel estimation using a small set of parameters and facilitate transmit/receive beamforming (BF) and spatial multiplexing. In radar systems, accurate signature extraction is crucial for target localization.

Traditionally, spatial delays at the array level are approximated as simple phase shifts under the narrowband assumption. However, this approximation breaks down when the instantaneous signal bandwidth becomes comparable to the reciprocal of the propagation delays across the array. Classical array processing addresses this with narrowband phase shift arrays \cite{van2004optimum} for the narrowband regime, and with true time delay (TTD) phase shifters \cite{hashemi2008integrated} for the wideband case. Nonetheless, TTD-based analog beamforming is challenging to scale to the high-dimensional signal spaces required in mmWave massive MIMO systems.

Until 5G, typical combinations of modest bandwidths and array sizes allowed spatial delay to be neglected, enabling conventional MIMO system designs. However, in mmWave and future ultra-massive MIMO systems, these delays become significant. Notably, at terahertz (THz) frequencies—candidates for 6G—even small arrays (e.g., $4\times4$) exhibit non-negligible spatial delays. As a result, different antenna elements may receive entirely different symbols, a phenomenon known as the delay-squint effect. Additionally, when wideband signals are processed through analog phase shifters, the hardware fails to maintain consistent angle and gain responses across the frequency range. This leads to the beam-squint effect, where different frequencies are effectively steered to different angles.

\subsection{Limitations on mmWave Massive-MIMO}

\subsubsection{Dual Wideband Effects}  
Addressing dual wideband effects—namely, space-dependent delays and frequency-dependent angles—is critical for both channel modeling and transceiver design in GHz-bandwidth ultra-massive MIMO systems. While the statistical modeling of these effects remains largely unexplored in the literature, a few recent works have proposed physical models that incorporate them \cite{brady2015wideband, wang2018spatial, Wang2019}. These physical models serve as a foundation for understanding the dual wideband phenomena, and emphasize the need to compensate for such effects when extending conventional spatial narrowband system models.

\subsubsection{Finite Sampling and Signal Processing Effects}  
An additional challenge arises from finite discrete sampling in the delay-space domain. When the input signal frequency is not an integer multiple of the sampling resolution—a common scenario—power leaks from the center of a frequency bin into adjacent bins \cite{lyons2004understanding}. Consequently, in discrete-time processing of mmWave massive MIMO systems, accurately identifying the continuous-domain signature of a scatterer or target becomes infeasible without addressing this leakage.

\subsubsection{Low SNR Scenario}  
At mmWave and terahertz (THz) frequencies, adverse propagation conditions can result in severely attenuated path gains, pushing signal-to-noise ratios (SNRs) as low as -15 dB \cite{akdeniz2014millimeter, liu2020deep}. Moreover, the impact of receiver-side impairments—such as thermal noise, phase noise, and amplifier noise—becomes more pronounced at higher operational frequencies \cite{haykin2001communication}. Several low-power wireless applications, including Wireless Sensor Networks (WSNs), the Internet of Things (IoT), and indoor positioning systems, also operate under similarly low SNR conditions. In such cases, ultra-low SNR severely limits the reliable extraction of spatio-temporal diversity branches, reducing the receiver's ability to exploit available diversity for robust performance.

\subsection{Related Work and Motivation}

The angle and delay components of the channel signature induced by scatterers in the wireless environment are estimated using fingerprint-based clustering in \cite{8307353}, which assumes a conventional spatial narrowband model for MIMO-OFDM systems. In \cite{brady2015wideband}, a line-of-sight (LoS) wideband model for Single Input Multiple Output (SIMO) systems is developed, where the dual wideband effect is quantified for the first time in the context of communication systems. Building upon this, \cite{wang2018spatial} presents a physical channel model for multipath, massive MIMO, and multi-carrier Orthogonal Frequency Division Multiplexing (OFDM) systems, alongside a channel estimation and user scheduling algorithm. In both \cite{wang2018spatial} and \cite{xie2019power}, the authors address power leakage effects primarily through rotation operations.

Further, \cite{wang2018spatial} and \cite{wang2019power} focus on modeling dual wideband and finite basis leakage effects but do not explicitly address the recovery of scatterer signatures embedded within ultra-low signal-to-noise ratio (SNR) environments. In contrast, \cite{Wang2019, jian2019angle} tackle spectral leakage due to grid mismatch using off-grid estimation methods—specifically, Sparse Bayesian Learning (SBL)—which are computationally intensive. Support Detection (SD)-based channel estimation approaches for dual wideband systems have also been proposed in \cite{gao2019wideband, tan2021wideband}, although these are primarily designed for lens antenna array configurations. Notably, none of the aforementioned studies provide a robust method for extracting channel scatterers obscured by severe receiver noise in ultra-low SNR scenarios.

In terms of learning-based approaches, \cite{he2018deep} introduces a denoising framework for lens antenna arrays without accounting for spatial wideband effects and without explicit estimation of angle-of-arrival (AoA) or time-of-arrival (ToA). Similarly, \cite{ye2020deep} demonstrates the utility of deep denoising networks, albeit for the purpose of channel state information (CSI) feedback rather than direct signature recovery. Liu et al. \cite{liu2020deep} propose a super-resolution compressive sensing method for Intelligent Reflecting Surfaces (IRS), incorporating significant sampling noise and enhancing channel estimation using a Denoising Convolutional Neural Network (DnCNN). A preliminary version of our work, focusing on deep learning (DL)-based denoising in a simplified setting, is presented in \cite{rai2022signature}.

To the best of the authors’ knowledge, no existing literature proposes a unified framework that combines deep learning-based channel denoising and clustering for the recovery of scatterers and fine-grained signature estimation under ultra-low SNR conditions in mmWave massive MIMO systems, while accounting for both dual wideband and finite basis leakage effects. Motivated by this gap, the present work introduces a novel end-to-end signature estimation framework that first leverages deep learning for path recovery in ultra-low SNR regimes, followed by an effective clustering mechanism for coarse-to-fine signature extraction.

\subsection{Contributions}

This paper presents a comprehensive study of system-level challenges associated with dual wideband effects, finite basis limitations, and high receiver noise in mmWave massive MIMO systems. The core problem of scatterer or target recovery and signature estimation is addressed using a Machine Learning (ML)-driven approach. The key contributions of this work are summarized as follows:

\begin{enumerate}
    \item The impact of finite basis constraints on dual wideband systems is analytically characterized using Fourier analysis, offering an accurate representation of dual wideband spreading in mmWave massive MIMO channels.
    
    \item A novel deep learning (DL)-based denoising framework is proposed for Channel Impulse Response (CIR) recovery, enabling robust scatterer extraction in ultra-low SNR scenarios.
    
    \item An unsupervised, apriori-free Local Gravitation-based Clustering (LGC) algorithm is developed to automatically identify the number of clusters and their spatial supports in sparse dual wideband environments.
    
    \item A fine-tuning algorithm is introduced to mitigate the finite basis effect and achieve super-resolution estimation of scatterer signatures.
    
    \item New performance metrics are proposed to rigorously evaluate the denoising and clustering components of the DL-ML framework, with extensive simulations validating its robustness and effectiveness.
\end{enumerate}

The rest of the paper is organized in the sections below. In section II, we first present a physical channel model for mmWave Massive MIMO system and discuss about the dual wideband, finite basis and ultra low SNR with high receiver noise effects. In section III, we give a brief background about image denoising, followed by robust data clustering in section IV. Further, in section V, we develop a coarse-fine signature estimation algorithm based on our proposed novel DL-ML approach. In section VI, we show the channel simulation results for various blocks involved in the framework. Finally, in section VII, we conclude the paper with future works.

\begin{table}[tbh]

\resizebox{\columnwidth}{!}{%
\begin{tabular}{ll}
\textbf{Notation}            & \textbf{Description}                                        \\
$(.)^T$                      & transposition of a matrix/vector                            \\
$(.)^H$                      & conjugate transposition of a matrix/vector                  \\
$(.)^*$                      & conjugation of a matrix/vector                              \\
$\circ$                      & elementwise product of two matrices                         \\
$E(.)$                       & expectation operator                                        \\
diag$\lbrace \bm{a} \rbrace$ & diagonal matrix as vector $\bm{a}$ in its main diagonal     \\
$\| \bm{a} \|_2$                       & l-2 norm of a vector $\bm{a}$                                        \\
diag$\lbrace \bm{a} \rbrace$ & diagonal matrix as vector $\bm{a}$ in its main diagonal     \\
$\|\bm{A}\|_F$               & Frobenius norm of a matrix $\bm{A}$                         \\
$[\bm{A}]_{i,j}$             & $(i,j)^{th}$ element of matrix $\bm{A}$                     \\
$(.)^+$                      & consider the value only if it is +ve else the value is zero
\end{tabular}%
}

\end{table}


\section{System Model}
\subsection{Dual Wideband Systems}
\begin{figure}
	\centering
	\includegraphics[width=1\linewidth]{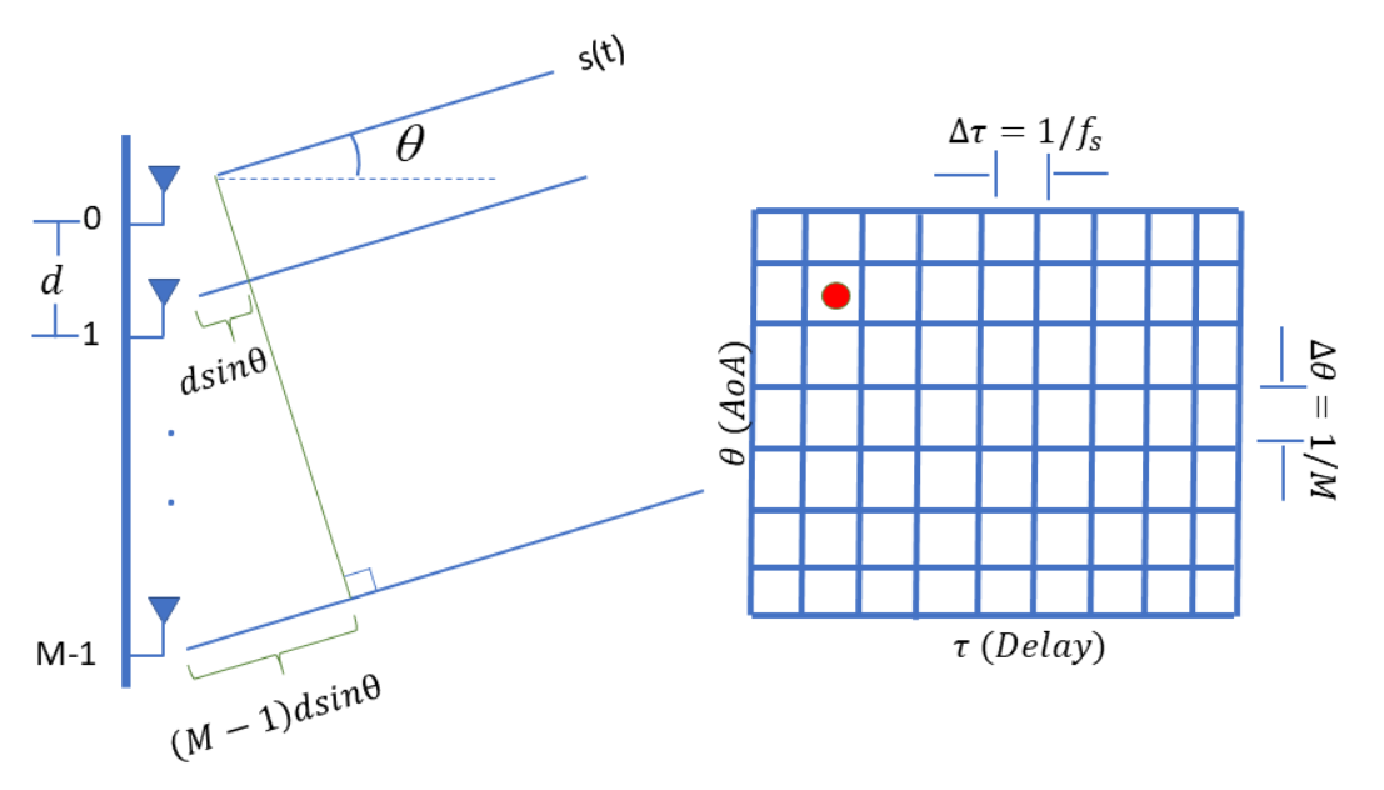}
	\caption{A ULA based system model for mmWave Massive MIMO system with illustration of channel signature.}
	\label{fig_systemmodel}
\end{figure}

We present the physical channel model for a mmWave massive MIMO system with Uniform Linear Array (ULA) geometry receiver as shown in Fig. \ref{fig_systemmodel}. We consider the array processing model for a wireless radio scene and describe the effect of different delays incurred during reception of a wideband signal at a large sized array. The baseband signal received at the $r^{th}(r=1 \cdots M)$ receive antenna from the $u^{th}$ user is

\begin{equation}
y_{u,r}(t)=\sum_{l=0}^{L_{u}-1}\widehat{\alpha}_{u,l}x_{u} \left(t-\tau_{u,l,r}\right) e^{-j2\pi f_{c}\tau_{u,l,r}}+n_{u,r}(t) 
\label{eq:system_model}
\end{equation}

where, $x_{u}(t)$ is the transmitted baseband signal of the $u^{th}$ user, $\widehat{\alpha}_{u,l}$ is the complex channel gain, $L_u$ are the number of physical paths faced by $u^{th}$ user, $\tau_{u,l,r}$ is the delay at $r^{th}$ receive antenna of the $u^{th}$ user via $l^{th}$ path and $n_{u,r}(t)$ is the Additive White Gaussian Noise (AWGN) following $\mathcal{CN}(0,\sigma^2)$. The delay $\tau_{u,l,r}$ is composed of three parts as shown below
\begin{equation}
\tau_{u,l,r}=\underbrace{\frac{r_{l}}{c}}_{part-I}+\underbrace{r\frac{d.\sin\phi_{u,l}}{c}}_{part-II} + \underbrace{\frac{v_{u}\cos\phi_{u,l}.t}{c}}_{part-III} .
\label{eq:delays}
\end{equation}
Part I in  \eqref{eq:delays} is  the delay due to the physical distance traveled by the multipath signal of the $u^{th}$ user to the $0^{th}$  receive antenna, giving rise to frequency selective nature of channel for wideband signals. Part II is the delay due to the extra distance traveled by the signal from $u^{th}$ user  for a scatter located at an angle $\phi_{u,l}$ through the aperture length of the $r^{th}$ antenna. This delay is non-negligible in comparison to symbol duration for a wideband signal (of BW GHz or more) when used with a massive antenna size (of 64 or more), giving rise to the spatial selective nature of the channel. Part III is the delay due to the mobility of the $u^{th}$ user for the underspread channel. For the static radio environment, we define, ${\lbrace\tau_{u,l},\theta_{u,l},L_u,\widehat{\alpha} \rbrace}$ as the signature of the $u^{th}$ user.

We can rewrite \eqref{eq:system_model}, ignoring the mobility effect as
\begin{equation}
y_{u,r}(t)=\sum_{l=0}^{L_{u}-1}\alpha_{u,l}x_{u}(t-\tau_{u,l}-r\frac{\theta_{u,l}}{f_{c}}) e^{-j2\pi r\theta_{u,l}} .
\end{equation}    

where, $\alpha_{u,l} \triangleq \widehat{\alpha}_{u,l}e^{-j2\pi f_c\tau_{u,l}}$ is effective complex channel gain and $\theta_{u,l} \triangleq \frac{d\sin\phi_{u,l}}{\lambda_{c}}$ is the spatial frequency.
Hence, we can write the uplink delay-spatial channel between $u^{th}$ user and $r^{th}$ antenna as

\begin{equation}
h_{u}(\tau,r)=\sum_{l=0}^{L_{u}-1}\alpha_{u,l} e^{-j2\pi r\theta_{u,l}}\delta\left(\tau-\tau_{u,l}-r \frac{\theta_{u,l}}{f_c}\right).
\end{equation}

After taking Fourier Transform (FT) of Eq. (4) w.r.t. $'\tau'$, we can write the frequency-spatial channel as
\begin{equation}
h_{u}(f,r)=\sum_{l=0}^{L_{u}-1}\alpha_{u,l}e^{-j2\pi r\theta_{u,l}}e^{-j2\pi f\tau_{u,l}}e^{-j2\pi fr\frac{\theta_{u,l}}{f_{c}}} .
\end{equation}

Now, using Multi Carrier Modulation (MCM) via Orthogonal Frequency Division Multiplexing (OFDM) with $N$ subcarriers of inter-carrier spacing $\Delta = f_s/N$, the channel response matrix for $u^{th}$ user in the subcarrier-antenna domain can be written as
\begin{equation}
\textbf H_{u}(n,r)=\begin{bmatrix}{h}_{u,0}(0)&\cdots&{h}_{u,0}((N-1)\Delta) \\{h}_{u,1}(0)&\cdots&{h}_{u,1}((N-1)\Delta)\\ \vdots\\{h}_{u,(R-1)}(0)&\cdots&{h}_{u,(R-1)}((N-1)\Delta) \end{bmatrix}
\end{equation}
\begin{equation}
\textbf{H}_{u}(n,r)=\sum_{l=0}^{L_{u}-1}\alpha_{u,l}\textbf{d}(\theta_{u,l})\textbf{c}^{T}(\tau_{u,l})\circ  \textbf{$\bm{S}$} (\theta_{u,l}) .
\label{eq:space_delay}
\end{equation}

Where,
$\textbf{d}(\theta_{u,l})\overset{\Delta}{=}[1,e^{-j2\pi\theta_{u,l}},.\quad.\quad.,e^{-j2\pi(M-1)\theta_{u,l}} ]^{T}$  is the $R \times 1$ direction vector (spatial beamforming vector), $\textbf{c}(\tau_{u,l})\overset{\Delta}{=}[1,e^{-j2\pi\Delta\tau_{u,l}},.\quad.\quad.,e^{-j2\pi(N-1)\Delta\tau_{u,l}} ]^{T}$ is the $N \times 1$ sub-carrier vector (temporal beamforming vector) and
$[\bm{S}_{u,l}]_{r,n}\triangleq exp(-j2\pi rn\Delta \frac{\theta_{u,l}}{f_{c}}) \quad \forall r,n$, the phase shift matrix with $\circ$ denoting the Hadamard product.

\subsection{Transform Domain-Advantage and Limitations}

We can transform the channel in \eqref{eq:space_delay} to its dual where we can observe various advantageous properties involved of the channel in delay-angle domain.
\begin{equation}
  \textbf{G}_u \triangleq \textbf{F}_{M}^H \textbf{H}_{u} \textbf{F}_{N}^{*},  
\end{equation}

where $\mathbf{F}_M$ denotes the $M\times M$ DFT matrix.

\begin{figure}[h]
	\centering	\includegraphics[width=0.95\linewidth]{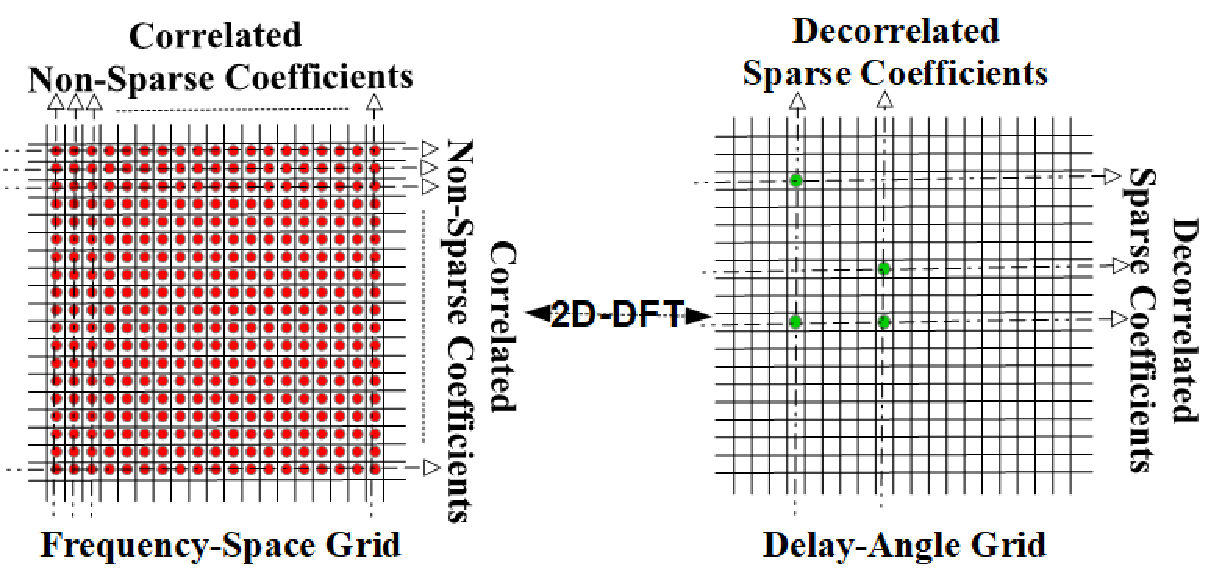}
	\caption{Correlated and deccorelated CIR structure in transform domains.}
	\label{fig_Corr_Decorr}
\end{figure}

Channel path sparsity and orthogonality in angle-delay domain is mathematically proven in \cite{wang2018spatial}. Figure \ref{fig_Corr_Decorr} visually illustrates that the channel coefficients exhibit non-sparsity and correlation in the space-frequency domain, while in the delay-angle domain, they become sparse and uncorrelated. However, the sparse and uncorrelated behaviour is restricted to the amount of leakage and the dual wideband spread.

\begin{figure}[h]
	\centering
	\includegraphics[width=0.80\linewidth]{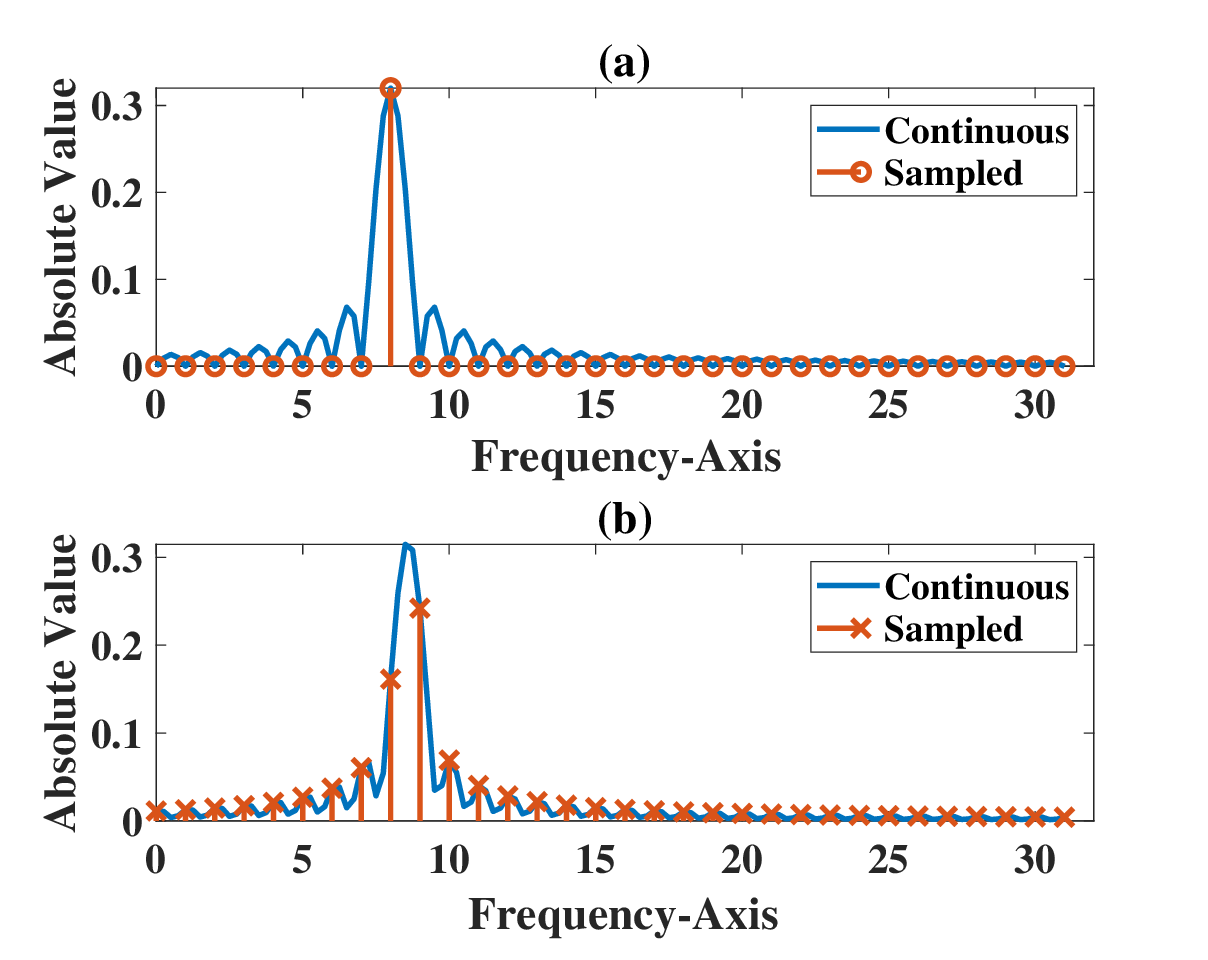}
	\caption{Finite basis effect and spectral leakage (a) Integer frequency (b) Non-Integer frequency.} 
	\label{fig_Diri1}
\end{figure}

Except for the frequencies that are integer multiples of the bin-resolution, leakage is inexorable for any general frequency to be estimated. In Fig. \ref{fig_Diri1}, it can be seen that for a signal with a single frequency of integer multiple of the bin, continuous magnitude spectrum samples only at a single point and zero elsewhere. On the other hand, in the case of non-integer multiple, the continuous magnitude spectrum is sampled at other points than the exact input frequency causing the leakage. 
\vspace{-0.5cm}

\begin{figure}[h]
	\centering
	\includegraphics[width=0.95\linewidth]{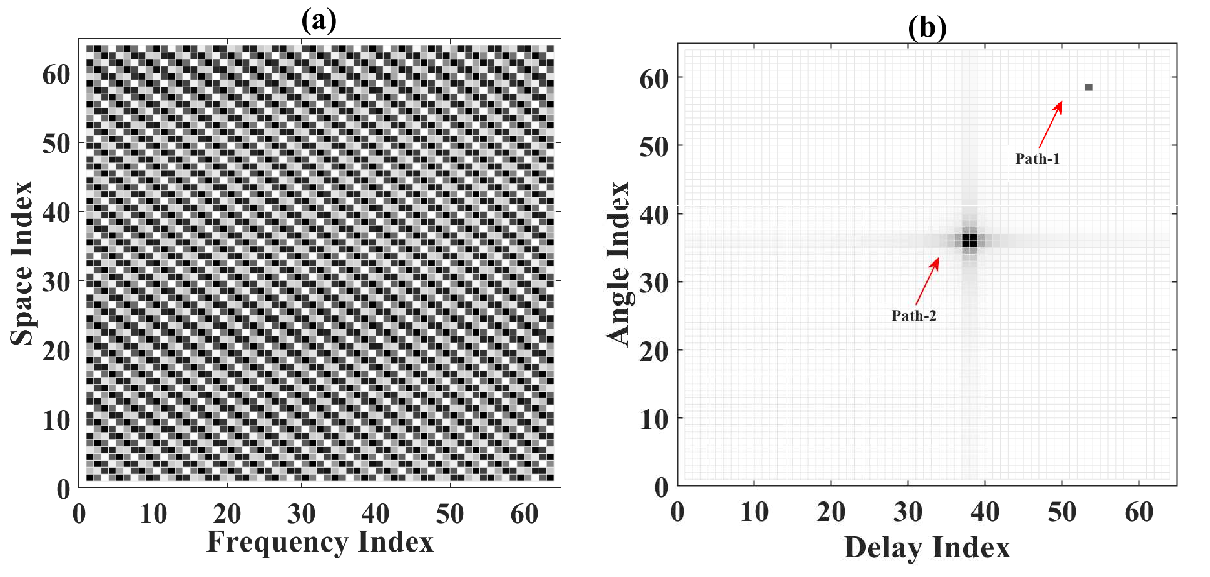}
	\caption{Dual Relationship (a) Space-Frequency domain (b) Angle-Delay domain.}
	\label{fig_image_sf_ad}
\end{figure}

A similar leakage effect is shown for the 2-D case in Fig. \ref{fig_image_sf_ad}. We observe that the \textit{Path-1} is an exact integer frequency multiple, whereas the \textit{Path-2} is a non-integer frequency multiple, causing the leakage in 2-D. Channel sparsity in angle-delay domain can also be observed from Fig. \ref{fig_image_sf_ad}(b).

\begin{figure}[h]
	\centering
	\includegraphics[width=1\linewidth]{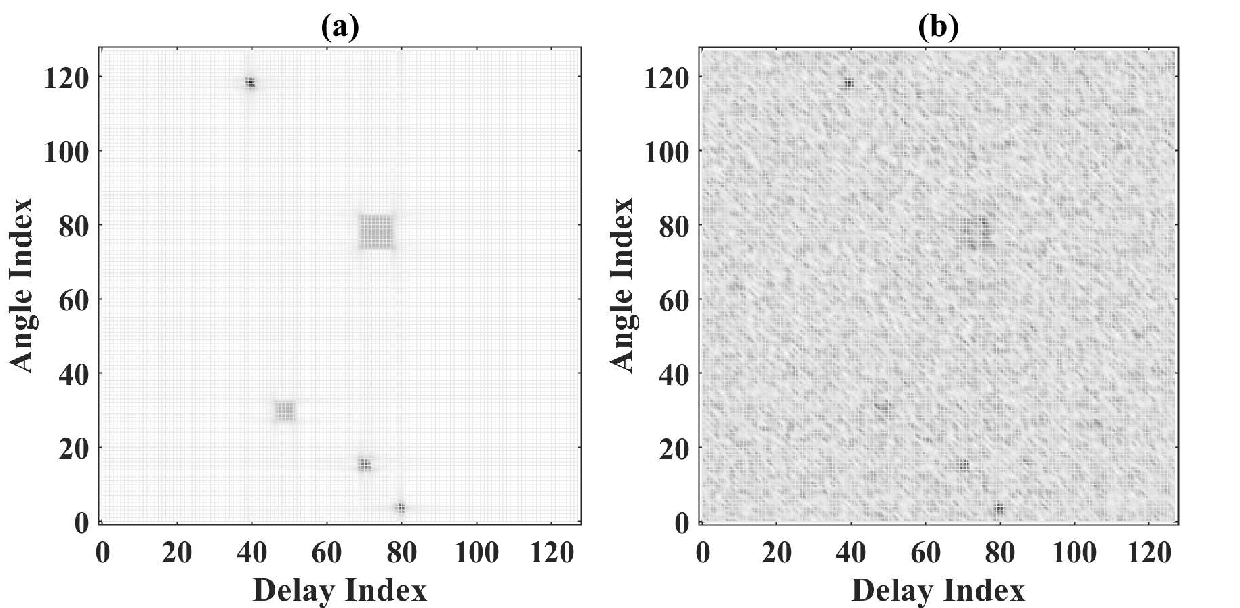}
	\caption{CIR for 5-physical paths (a) Delay-Angle spreading Effect (b) Noise effect at -15dB SNR.}
	\label{fig_DWB}
\end{figure}

Furthermore, due to the dual wideband spread, there is a 2-D delay-angle spreading of the channel paths as shown in Fig. \ref{fig_DWB}(a) and the CIR exhibits block sparsity. It can be noted from Fig. \ref{fig_DWB}(b) the effect of ultra-high noise ($\geq -15 dB$) suppresses the channel signatures drastically.

\vspace{-0.32cm}
\begin{figure*}[h]
	\centering
	\includegraphics[width=0.8\linewidth]{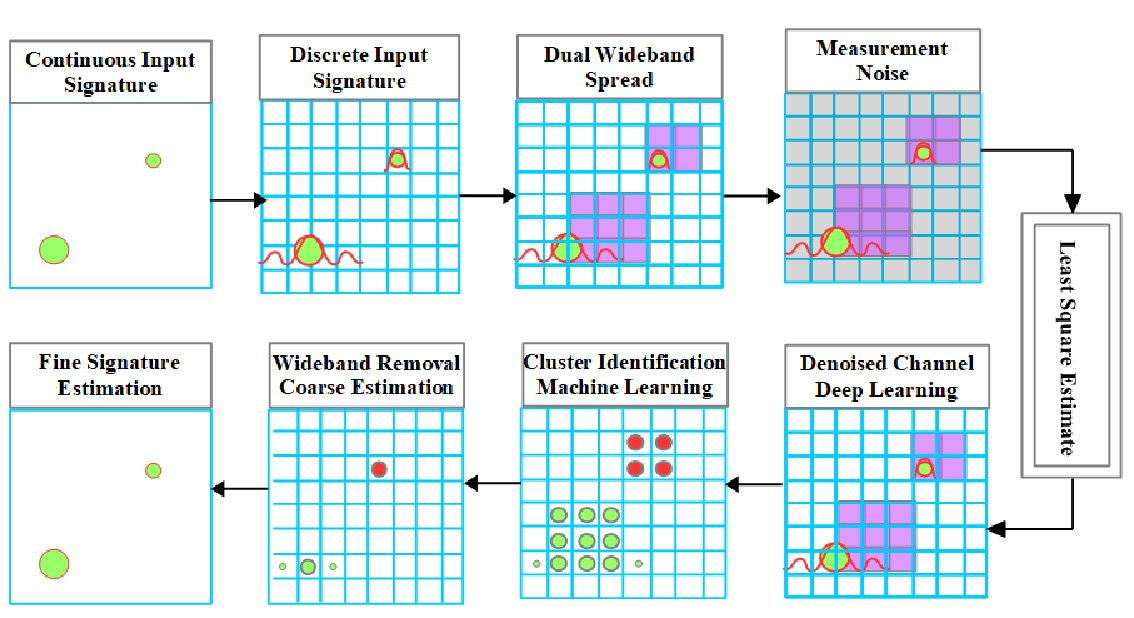}
	\caption{Signal processing engine for mmWave Massive MIMO systems considering dual wideband and finite basis effects.}
	\label{fig_sig_flow}
\end{figure*}
\vspace{0.1cm}

 \subsection{Proposed AI-Enabled Framework}

We summarize the key challenges and present an end-to-end AI-enabled signal processing framework for signature estimation in dual wideband mmWave massive MIMO systems, as illustrated in Fig. \ref{fig_sig_flow}. The continuous-time channel signatures, when projected onto a finite space-time grid, are affected by finite basis spreading, followed by the dual wideband effect. Additionally, the receiver operates under high measurement noise, further complicating channel estimation. To recover and estimate the channel paths, we begin with an initial preamble-based channel estimate using the Least Squares (LS) method. However, under ultra-high receiver noise conditions, the channel path components cannot be directly identified. To address this, we apply a deep learning (DL)-based denoising technique to the delay-angle domain channel response, followed by a robust, unsupervised clustering algorithm that does not require prior knowledge of the number of paths. This clustering step provides a coarse signature estimate and helps mitigate the wideband effects. Finally, a fine-tuning mechanism is employed to address residual spectral leakage, refining the signature estimates to achieve high-resolution path recovery.

\section{Image Denoising: Background}

Once the channel response is obtained in the delay-angle domain, it is significantly corrupted by noise in ultra-low SNR scenarios. To enable accurate identification of channel signatures, the first essential step is to suppress this observation noise as effectively as possible. Notably, the CIR in the delay-angle domain exhibits the structural properties of a natural two-dimensional (2D) image. As a result, the denoising of the noisy CIR matrix can be formulated as an image denoising problem.

Image denoising is a well-established task in the image processing literature, with numerous frameworks available for recovering clean images from noisy observations. A major advancement in this field has been the adoption of Deep Neural Networks (DNNs), which offer powerful representation capabilities by extracting hierarchical features through a sequence of nonlinear transformations. In a DNN, the output features are produced through a cascade of nonlinear functions, ultimately mapping to a space where they are linearly separable \cite{hornik1989multilayer}. For a given input feature vector $\mathbf{x}$, the output can be represented as:
\setlength{\abovedisplayskip}{3pt}
\begin{align}
\text{Output} = f^{(k)}(f^{(k-1)}(...f^{(2)}(f^{(1)}(\mathbf{x}))...)).
\end{align}

However, deep networks often encounter practical challenges, such as vanishing gradients in very deep architectures, slow convergence, and sensitivity to parameter initialization. To mitigate these issues, two widely adopted techniques—**Residual Learning** \cite{he2016deep} and **Batch Normalization (BN)** \cite{ioffe2015batch}—have significantly improved the stability and performance of deep learning models.

In this work, we adopt the pre-trained Denoising Convolutional Neural Network (DnCNN) proposed in \cite{7839189} for CIR denoising. The choice is motivated by the following benefits:
\begin{itemize}
	\item Its deep architecture effectively captures image-like features of the CIR matrix.
	\item It supports blind Gaussian denoising across a range of noise levels.
	\item It performs well on sparse images, making it suitable for wireless channels with few dominant paths.
\end{itemize}

For a clean image vector $\mathbf{x}$ and additive noise vector $\mathbf{n}$, the noisy observation is modeled as $\mathbf{y} = \mathbf{x} + \mathbf{n}$. In traditional image denoising, the goal is to learn a mapping $\mathcal{F}(\mathbf{y}) = \mathbf{x}$ to recover the clean image. In DnCNN, however, a **residual learning** approach is adopted: the network learns the noise component as $\mathcal{R}(\mathbf{y}) = \mathbf{n}$, from which the clean image is then obtained as $\mathbf{x} = \mathbf{y} - \mathcal{R}(\mathbf{y})$. Given a training set of $N$ noisy-clean image pairs $\lbrace(\mathbf{y}_i, \mathbf{x}_i)\rbrace_{i=1}^N$, the network is trained to minimize the mean squared error between the predicted and actual noise:
\begin{align}
J(\mathbf{\Theta}) = \frac{1}{2N} \sum_{i = 1}^{N} \parallel \mathcal{R}(\mathbf{y}_i; \mathbf{\Theta}) - (\mathbf{y}_i - \mathbf{x}_i) \parallel^2,
\end{align}
where $\mathbf{\Theta}$ represents the network parameters.

We use these pre-trained parameters $\mathbf{\Theta}$ and demonstrate the efficacy of deep networks for denoising CIR matrices in challenging ultra-low SNR conditions. For further architectural and hyperparameter details, readers are referred to \cite{7839189}. More importantly, we introduce a performance metric specifically designed to assess denoising effectiveness for wireless scenarios. If the channel path powers are not normalized to unity, then, for a fixed receiver noise variance, the SNR of the channel response differs from its equivalent image-domain SNR.  
\footnote{When the Discrete Fourier Transform (DFT) matrix columns are not normalized, power distribution changes in the transform domain, and we obtain DFT processing gain, enabling operation in lower SNR regimes.}

To quantify denoising performance, we define the Image-Equivalent SNR for the wireless scenario both Before Denoising (BD) and After Denoising (AD). This is computed by referencing a noise-free CIR against the noisy and denoised CIRs.

\begin{equation}
\begin{aligned}
    \gamma_{{\textstyle\mathstrut}BD }  \overset{\Delta}{=}   \frac{{\parallel \mathbf{\widehat{G}_{C}} \parallel}^2_F}{{\parallel {\mathbf{\widehat{G}}_{N}-\mathbf{\widehat{G}_{C}}} \parallel}^2_F} \hspace{0.2cm} , \hspace{0.2cm} 
    \gamma_{{\textstyle\mathstrut}AD }  \overset{\Delta}{=}   \frac{{\parallel \mathbf{\widehat{G}_{C}} \parallel}^2_F}{{\parallel {\mathbf{\widehat{G}}_{D}-\mathbf{\widehat{G}_{C}}}_F \parallel}^2_F} .
\end{aligned}
\end{equation}

\vspace{0.2cm}

\section{Robust Data Clustering: Background}
 In order to find the no. of paths and initial (coarse) channel signature estimates, we need a robust data clustering algorithm without having any apriori knowledge about source scatters and background noise statistics. Unlike the k-means clustering algorithm, we have to use a clustering algorithm that gives us the cluster outputs to identify the number of scatters present in the channel, in addition to the support of cluster spreads. In this work, we develop the framework of Local Gravitation based Clustering (LGC) \cite{7915751} for finding the no. of scatters and their initial signatures. The key reasons to use LGC is as below

\begin{itemize}
	\item By using the concept of Local Resultant Forces (LRFs) of neighboring points in LGC, the relation existing between the points and their neighbors is exploited.  
	\item Similar to density-based approaches - it is able to identify the no. of clusters and the non-spherical shapes but with lesser complexity.
\end{itemize}

However, we can also use the k-means algorithm by appropriately applying a mechanism for picking the optimal number of unknown clusters. As an example, we can choose \textit{elbow method} for identifying the number of clusters. Nevertheless, the k-means opted with the elbow method needs an upper bound on the number of clusters-k, which can be leading to erroneous results in case of ultra-low SNR and improper thresholding.

Therefore, the final outcome of the data clustering algorithm applied to a noisy dataset is influenced by three key factors: $i)$ \textit{the effectiveness of the denoising process}, $ii)$ \textit{the accuracy of the hard thresholding}, and $iii)$ \textit{the choice of the clustering algorithm}.
Once we denoise the dataset using the DL method, we put a hard threshold to further clean the denoised response clustering dataset, which directly impacts the clustering performance as shown in the next section.

\subsubsection*{Hard Thresholding} 
We define two types of hard thresholding techniques to truncate the dataset before clustering.

\begin{enumerate}
    \item \textit{Energy Threshold (ET):} The energy threshold removes low-energy components based on a global cutoff. The truncated dataset is defined as:

    \begin{equation}
        \mathbf{\dot{G}}_{ET} = \Big \lbrace 
        (i,j) : [\mathbf{\hat{G}}_{D}]_{i,j} > \sqrt{{\parallel \mathbf{\hat{G}}_{D} \parallel}^2_F / NM}  \Big \rbrace,
        \label{eq:energy_threshold}
    \end{equation}
    where $\parallel \cdot \parallel_F$ denotes the Frobenius norm, and $N, M$ are the dimensions of $\mathbf{\hat{G}}_D$.

    \item \textit{Percentile Threshold (PT):} The percentile threshold removes values below a specified $\alpha$-percentile. The truncated dataset is given by:
    
    \begin{equation}
        \mathbf{\dot{G}}_{PT} = \Big \lbrace 
        (i,j) : [\mathbf{\hat{G}}_{D}]_{i,j} > \mathcal{P_{\alpha}}(\mathbf{\hat{G}}_{D}) \Big\rbrace,
        \label{eq:percentile_threshold}
    \end{equation}
     where $\mathcal{P_{\alpha}}(\cdot)$ denotes the $\alpha$-percentile function. Here, $\alpha$ is a sparsity parameter (e.g., $\alpha=95$ retains the top 5\% of values) and must be selected a priori based on the desired level of sparsity in the dataset.
\end{enumerate}

In order to evaluate the quality of the clustered output, we define a clustering performance metric based on the spatial compactness of each identified cluster. Specifically, for the $i^{\text{th}}$ cluster, we compute the standard deviations of its support in the delay and angle domains, denoted by $\sigma_d^i$ and $\sigma_a^i$, respectively. These quantities reflect the spread of each cluster along the two dimensions.

The overall clustering metric (CM) is then defined as the product of the two-dimensional spreads of all detected clusters:

\begin{equation}
    CM = \prod_{i=1}^{\hat{L}} \sqrt{ \left( \sigma_d^i \right)^2 + \left( \sigma_a^i \right)^2 },
\end{equation}

where $\hat{L}$ is the total number of detected clusters. A lower CM value indicates tighter (i.e., more compact) clusters, which is desirable for accurate scatter/path characterization.

We also define the Mean Absolute Error (MAE) as a metric to evaluate the accuracy of the estimated number of clusters produced by the clustering algorithm. It is given by:

\begin{equation}
    \text{MAE}(L) = \mathbb{E} \left\lbrace \left| \hat{L} - L \right| \right\rbrace,
\end{equation}

where $L$ is the true number of clusters and $\hat{L}$ is the estimated number. A lower MAE indicates a more accurate estimation of the true cluster count.

To ensure robust clustering of the noisy CIR, both the CM — which quantifies the compactness of clusters — and the MAE should be minimized. To jointly capture both aspects of clustering quality, we introduce the Effective Clustering Metric (ECM), defined as:

\begin{equation}
    \text{ECM} \triangleq \mathbb{E} \left\lbrace \left(1 + \text{MAE} \right) \cdot \log_{10}(\text{CM}) \right\rbrace.
\end{equation}

A lower ECM value indicates that the clustering algorithm is both accurate in estimating the number of clusters and effective in forming compact clusters.

Finally, it is important to note that overall clustering performance is also influenced by the quality of denoising. The denoising and clustering stages are interdependent, and both jointly affect the accuracy of the estimated path Direction of Arrival (DoA) and Time of Arrival (ToA) signatures.

\section{Coarse-to-Fine Signature Estimation}

In this section, we present the proposed signature estimation algorithm, detailed in Algorithm~\ref{algo:alg_1}, for dual wideband systems. The algorithm operates under the assumptions that (i) the peak of each cluster remains aligned with its corresponding coarse Angle of Arrival (AoA) and Time of Arrival (ToA) bins, and (ii) there is no overlap between the signatures of distinct physical scatterers.

The objective is to estimate the signature of a physical scatterer, which may not lie exactly on integer multiples of the sampling resolution in either the delay or angle domain. To achieve this, we first perform a coarse signature extraction to identify the initial support regions of the scatterers. This is followed by a fine-tuning step, which refines the signature estimates by accounting for dual wideband dispersion and the effects of a finite basis resolution.

\begin{algorithm}[h]

\small
	\KwIn{$\mathbf {y_{u}}$ : Received signal vector  }
	\KwOut{$\hat{L}_{u}, \hat{\tau}_{u,l}, \hat{\theta}_{u,l}$}
	Find conventional LS channel estimate-$\bm{\hat{H}}_u$ \\
	Take 2-D IDFT to convert $\bm{\hat{H}}_u$ to Delay-Angle domain $\bm{\widehat{G}}$ \\
	Denoise channel image (normalize CIR) using DnCNN\\
	Normalize back the denoised channel-image equivalent to CIR $\bm{\widehat{G}}_{D}$\\
    Put energy/percentile threshold to prepare cluster dataset-$\bm{\dot{G}}$\\
    Implement LGC clustering algorithm\\
    output:  No. of Clusters $\hat{L}_{u}$, Cluster spreads in Delay-Angle $\bm{\ddot{G}}_{u,l}$\\
	\For{$ l \leftarrow 1 $ \KwTo $ \hat{L}_{u} $}
		{
		
		Select the dominant value of cluster as coarse estimation from \eqref{eq_coarse}\\
            
        Remove wideband effect via conjugation and apply the rotation to estimate the fine-tuned signature using \eqref{eq_rem_dwb}-\eqref{eq_final}.
 			}
	\caption {Spatio-Temporal Signature Estimation} \label {algo:alg_1}
\end{algorithm}

In the preamble phase, we first obtain a coarse estimate of the channel impulse response (CIR) in the spatial-frequency domain, denoted by $\mathbf{\widehat{H}}$, using the conventional Least Squares (LS) method \cite{barhumi2003optimal}. Following this, a 2D inverse discrete Fourier transform (IDFT) is applied to $\mathbf{\widehat{H}}$ to convert the estimated CIR into the angular-delay domain, yielding $\mathbf{\widehat{G}}$.

In scenarios with low noise, the number of clusters in $\mathbf{\widehat{G}}$ can be directly identified. However, under high noise conditions, the signatures of interest may be buried beneath the noise floor and lost during simple hard thresholding. To address this, we leverage DL-based denoising techniques rooted in image processing. Specifically, we normalize the CIR to form its image-equivalent representation and feed it into the DnCNN denoiser \cite{7839189}, a deep convolutional neural network pre-trained on natural image datasets. Once the denoising is complete, the output is re-normalized back to the CIR scale, and an energy or percentile threshold is applied to obtain a cleaner dataset for clustering. This thresholded, denoised CIR is denoted by $\bm{\dot{G}}_{u}$.

Next, we apply the robust LGC algorithm to the denoised 2D spatial dataset $\bm{\dot{G}}_{u}$. This process yields an estimated number of clusters $\hat{L}$, along with their respective support regions, denoted by $\bm{\ddot{G}}_{u,l}$. From each cluster, we determine the coarse signature bin as

\begin{equation}
    (m_l,n_l) = \underset{(m,n)}{\operatorname{argmax}} \lbrace\bm{\ddot{G}}_{u,l}\rbrace \quad \forall \quad l \in \{1, 2, \dots, \hat{L}\}.
    \label{eq_coarse}
\end{equation}

To mitigate power leakage effects and obtain a more accurate estimate, we apply a 2D rotation-based refinement technique. Around each coarse bin $(m, n)$, we define a fine-resolution grid point $(\delta_m, \delta_n) \in [-\frac{\pi}{M}, \frac{\pi}{M}] \times [-\frac{\pi}{N}, \frac{\pi}{N}]$. The corresponding rotation matrices are defined as

\begin{equation}
\begin{aligned}
    \bm{F}_{r}^{m}(\delta_m) & \triangleq \text{diag} \lbrace 1, e^{j2\pi\delta_m}, \cdots, e^{j2\pi(M-1)\delta_m} \rbrace, \\
    \bm{F}_{r}^{n}(\delta_n) & \triangleq \text{diag} \lbrace 1, e^{j2\pi\delta_n}, \cdots, e^{j2\pi(N-1)\delta_n} \rbrace.
\end{aligned}
\end{equation}

Due to the dual wideband nature of the signal model (as defined in Eq.~\eqref{eq:space_delay}), we compensate for the associated spread around the coarse bin using

\begin{equation}
    \bm{\tilde{H}}_{u,l}^{r} = \bm{\hat{H}}_{u,l} \circ \bm{S}^{*}(\hat{\theta}_{u,l}),
    \label{eq_rem_dwb}
\end{equation}

where $\bm{S}^{*}(\cdot)$ models the dual wideband effect and $\circ$ denotes element-wise multiplication. The fine rotation-based refinement is then performed via:

\begin{equation}
    (\hat{\delta}_m, \hat{\delta}_n) = \underset{(\delta_m, \delta_n)}{\operatorname{argmax}} \quad \| \bm{f}^{H}_{m,l} \bm{F}_{r}^{m}(\delta_m) \bm{\tilde{H}}_{u,l}^{r} \bm{F}_{r}^{n}(\delta_n) \bm{f}^{*}_{n,l} \|_2^2.
    \label{eq_rot_max}
\end{equation}

Finally, the fine-tuned \footnote{ Modulo operations inherent in DFT-based calculations are omitted here for clarity but must be handled properly during implementation, especially when dealing with negative values.} estimate of the signature is computed as:

\begin{equation}
\begin{aligned}
    (m^f, n^f) &= (m + \hat{\delta}_m, n + \hat{\delta}_n), \\
    (\hat{\theta}_{u,l}, \hat{\tau}_{u,l}) &= \left( \frac{m^f}{M}, \frac{n^f}{N\Delta} \right).
    \label{eq_final}
\end{aligned}
\end{equation}

We introduce the Denoised Mean Square Error (DMSE) as a novel evaluation metric to rigorously assess the impact of denoising and clustering on the accuracy of fine-tuned angular-delay signature estimates. Since noisy observations can lead to false detections or missed scatterers, the DMSE metric focuses solely on the falsely detected clusters.

\begin{equation}
    \text{DMSE} \triangleq \mathbb{E} \left\lbrace \frac{\text{NMSE}_{(\theta,\tau)}}{1 + N_F} \right\rbrace,
    \label{eq_DMSE}
\end{equation}

where the normalized mean square error over direction and delay is defined as

\begin{equation}
    \text{NMSE}_{(\theta,\tau)} \triangleq \frac{1}{\hat{\Tilde{L}}} \sum_{l=1}^{\hat{\Tilde{L}}} \left( \frac{\| \hat{\theta}_l - \theta_l \|_2^2}{\| \theta_l \|_2^2} + \frac{\| \hat{\tau}_l - \tau_l \|_2^2}{\| \tau_l \|_2^2} \right),
\end{equation}

and the number of false detections is

\begin{equation}
    N_F \triangleq (\hat{L} - L)^+.
\end{equation}

 \section{Simulation Results}
 In this section, we present a dual wideband channel realization, followed by the corresponding signature estimation procedure. An uplink channel is synthesized at a carrier frequency of 58~GHz, incorporating four distinct physical propagation paths. Each path is generated based on a delay spread of 15~ns, as specified in \cite{CM}. The angles of arrival are independently drawn from a uniform distribution, $\theta \sim \mathcal{U}(-\frac{\pi}{2}, \frac{\pi}{2})$, and the complex channel coefficients are sampled from a Rayleigh distribution. The complete set of simulation parameters is summarized in Table~\ref{tab:sim_para}.

 \begin{table}[h]
 	\centering
 	\caption{Simulation Parameters}
 	\label{tab:sim_para}
 	\small\addtolength{\tabcolsep}{-5pt}
 	\begin{tabular}{|c|c|c|}
 		\hline
 		\textbf{Sl.No.} & \textbf{Specification}   & \textbf{Amount} \\ \hline \hline
 		
 		1.&Uplink Carrier Frequency  & 58 GHz  \\ \hline
 		2.&No. of Subcarriers(N) &128  \\ \hline
 		3.&No. of Antennas (M) & 128 \\ \hline
 		4.&System bandwidth $(f_{s}$) & (0.1-0.2)$f_{c}$  \\ \hline
 		5.&Delay Spread & 15ns \cite{CM} \\ \hline
 		6.&No. of Physical Channel Paths & 2-4  \\ \hline
 		7.&Modulation Scheme& QPSK  \\ \hline	
 		8. & Angle of Arrival & $\mathcal{U} \tilde{(- \pi/2 , \pi/2)} $ \\ \hline
 		9. & Time of Arrival & $ exp (0,\tau_{max})$ \\ \hline
 	\end{tabular} 
 \end{table}
 
\begin{table*}[t]
 \caption{Performance of Denoising in Terms of Image Equivalent-SNR.}
\label{Table:Table_denoising}
\begin{center}
\begin{tabular}{| M{2cm} | M{2cm} | c | c | c |  c | c | c |}
\hline
\textbf{Rx SNR} & \textbf{SNR BD } & \multicolumn{3}{ c |}{\textbf{SNR AD}}  & \multicolumn{3}{ c |}{\textbf{Relative SNR }} \\ 
 \cline{3-5} \cline{6-8} 
 &  & \textbf{Mean} & \textbf{Median} & \textbf{DnCNN}  & \textbf{Mean} & \textbf{Median} & \textbf{DnCNN} \\
 \hline

-25   & -15.4887 & -14.5541 & -14.5541 & -14.4088   & 6.03\% & 6.03\% & 6.97\% \\ \hline
-20   & -10.3524 & -9.4430 & -9.2267 & -9.1974   & 8.78\% & 10.87\% & 11.15\% \\ \hline
-15   & -5.1067 & -4.3723 & -4.1935 & -3.8445   & 14.38\% & 17.88\% & 24.71 \% \\ \hline
-10   & 0.1787 & 0.3205 & 0.4159 & 1.2781   & 79.35 \% & 132.73 \% & 615.22 \% \\ \hline
-5   & 5.4594 & 3.9889 & 3.9168 & 5.9343   & -26.93 \% & -28.25 \% & 8.69 \% \\ \hline
0   & 10.7281 & 6.1528 & 5.9272 & 10.1970 & -42.64 \% & -44.75 \% & -4.95 \% \\ \hline

\end{tabular}
 
\end{center}
\end{table*}

 \begin{figure}[h]
	\centering
	\includegraphics[width=0.80\linewidth]{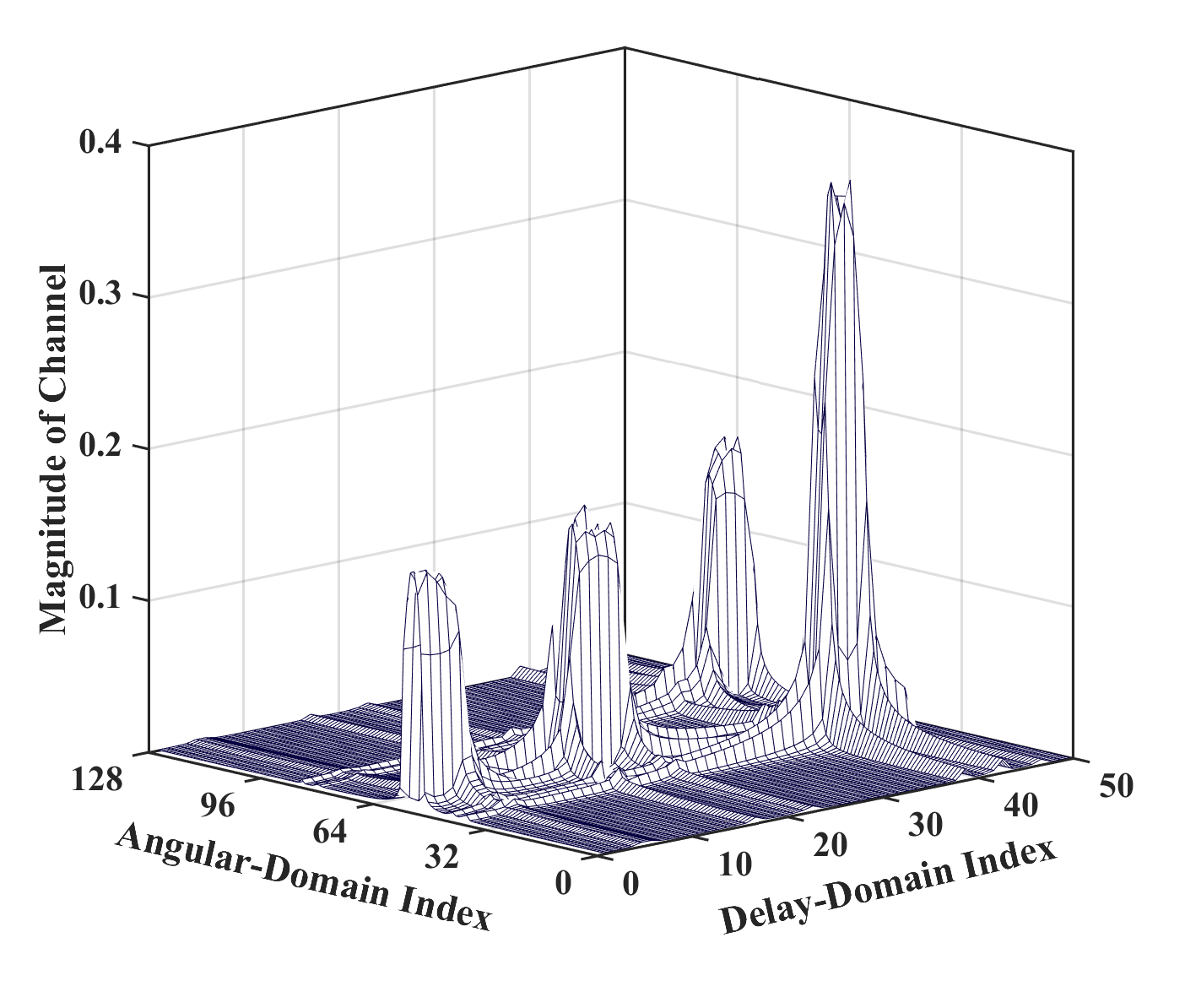}
	\caption{Channel realization with 4-physical paths at $f_c=58$GHz, $f_s=0.2f_c$.}
	\label{fig_chan_ip}
\end{figure} 
   
\begin{figure}[h]
	\centering
	\includegraphics[width=0.80\linewidth]{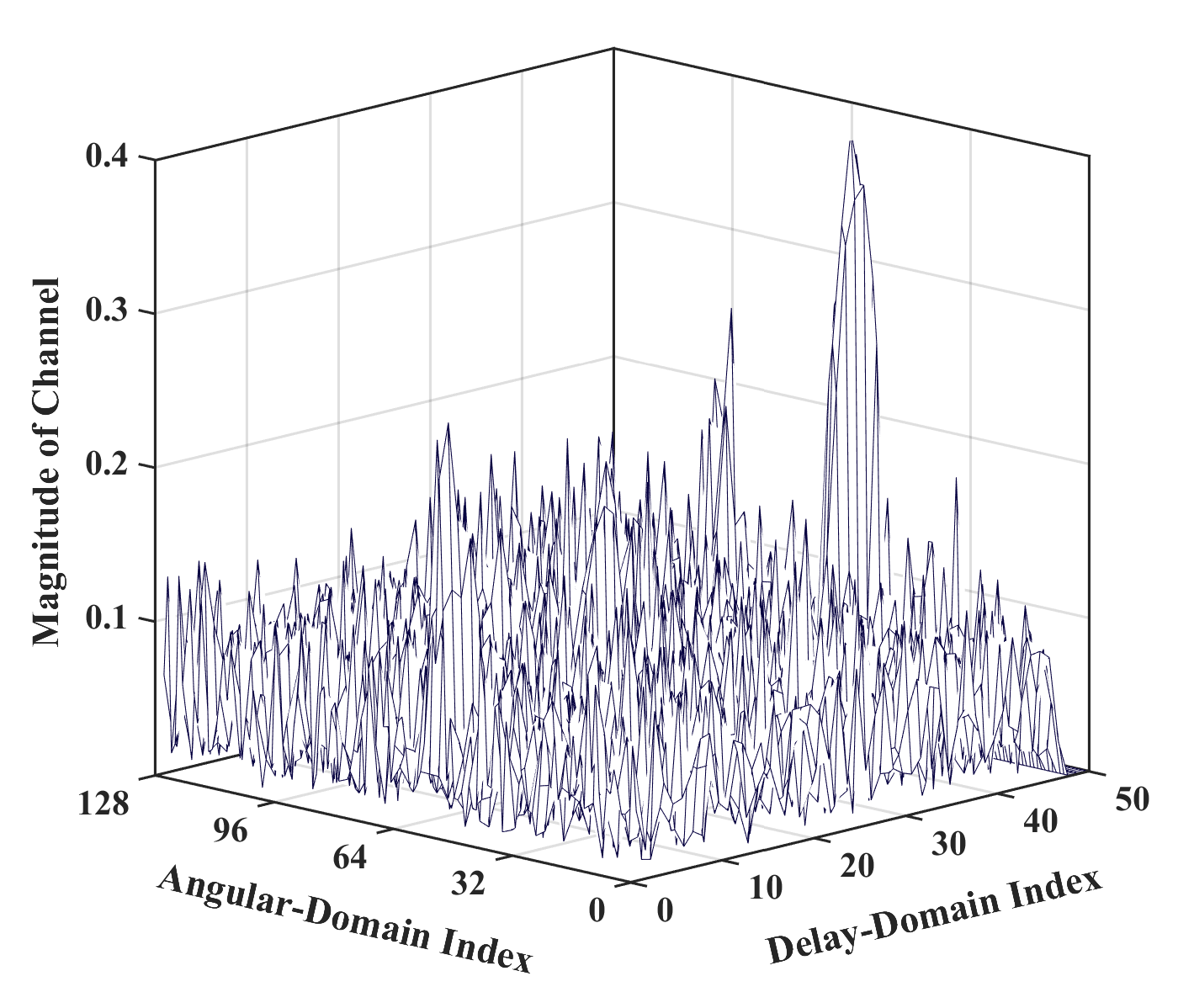}
	\caption{LS Estimate Output of CIR in Delay-Angle at $-15$dB.}
	\label{fig_chan_ls}
\end{figure}

 The CIR for four physical paths is illustrated in Fig.~\ref{fig_chan_ip}. The figure clearly reveals the two-dimensional dual wideband spread effect, which becomes prominent due to a high bandwidth-to-carrier-frequency ratio (approximately 0.2). Additionally, the ridge-like patterns observed are attributed to the finite number of antennas ($M$) and subcarriers ($N$), illustrating leakage effects caused by the finite basis representation.

At the receiver, assuming a reception SNR of $-15$\,dB, we first perform the LS estimation of the CIR. After transforming the spatial-frequency domain CIR to the delay-angle domain, as shown in Fig.~\ref{fig_chan_ls}, it becomes evident that most of the physical paths are submerged in noise. Consequently, applying either an arbitrary threshold or a threshold based on the average noise floor is insufficient for recovering all the paths.

To address this challenge, we employ the proposed denoising and clustering framework described in Algorithm~\ref{algo:alg_1}, which enables more effective path recovery under noisy conditions.

\begin{figure*}[h]
	\centering
	\includegraphics[width=0.8\linewidth]{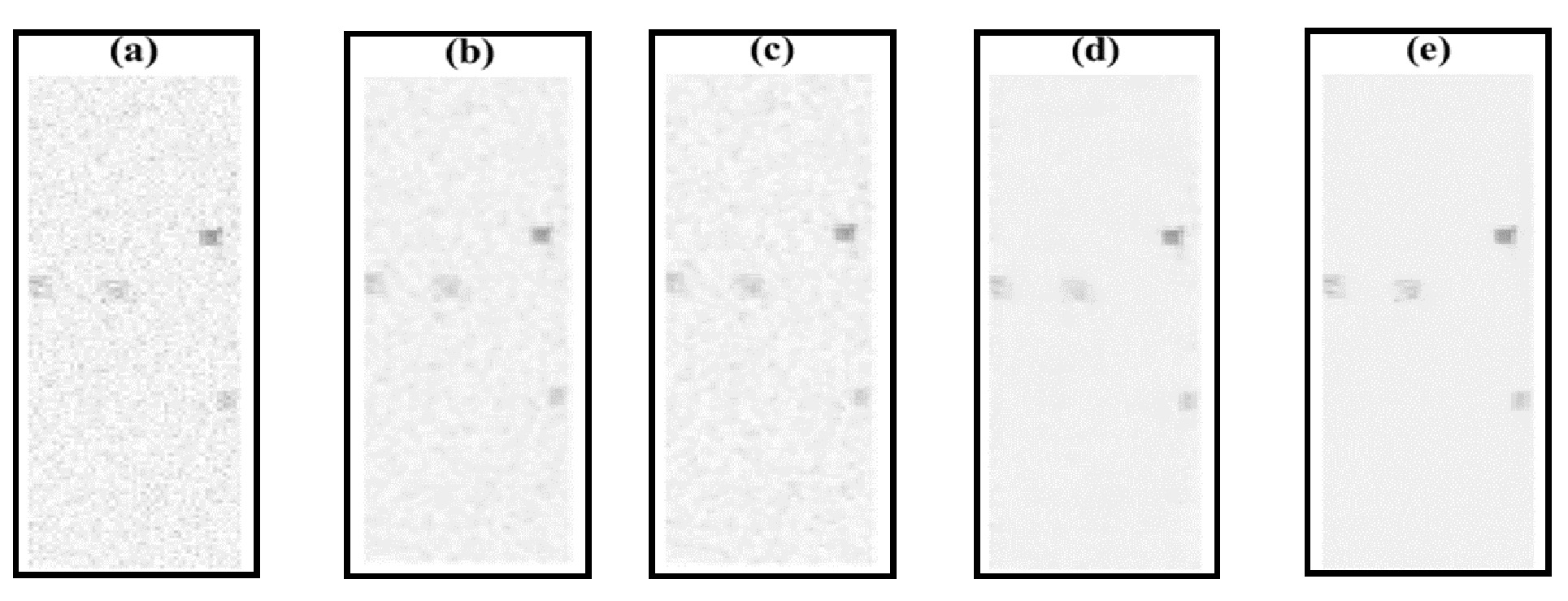}
	\caption{Image equivalent of CIR (a) Noisy image at -15dB SNR (b) Denoised image using mean filtering (c) median filtering (d) DnCNN with unknown noise variance (e) DnCNN with known variance.}
	\label{fig_chan_denoising}
\end{figure*}

 \subsection{Performance of Denoising:} 
 
 We begin by normalizing the wireless CIR to the range $[0,1]$ to form its image-equivalent representation, as shown in Fig.~\ref{fig_chan_denoising}(a). At an SNR of $-15$\,dB, the individual paths are indistinguishable from noise in this representation. To assess traditional denoising techniques, we apply mean and median filters to the normalized CIR image. The corresponding outputs are shown in Figs.~\ref{fig_chan_denoising}(b) and \ref{fig_chan_denoising}(c), respectively. However, these methods yield limited visual improvement, with path signatures still largely obscured. In contrast, Fig.~\ref{fig_chan_denoising}(d) demonstrates the result of denoising using the DnCNN model, assuming an unknown noise variance. This approach successfully reveals four distinct patches, corresponding to the underlying physical paths. The restoration quality improves further when a known noise level map is provided, as illustrated in Fig.~\ref{fig_chan_denoising}(e).

\begin{figure}[h]
	\centering
	\includegraphics[width=0.85\linewidth]{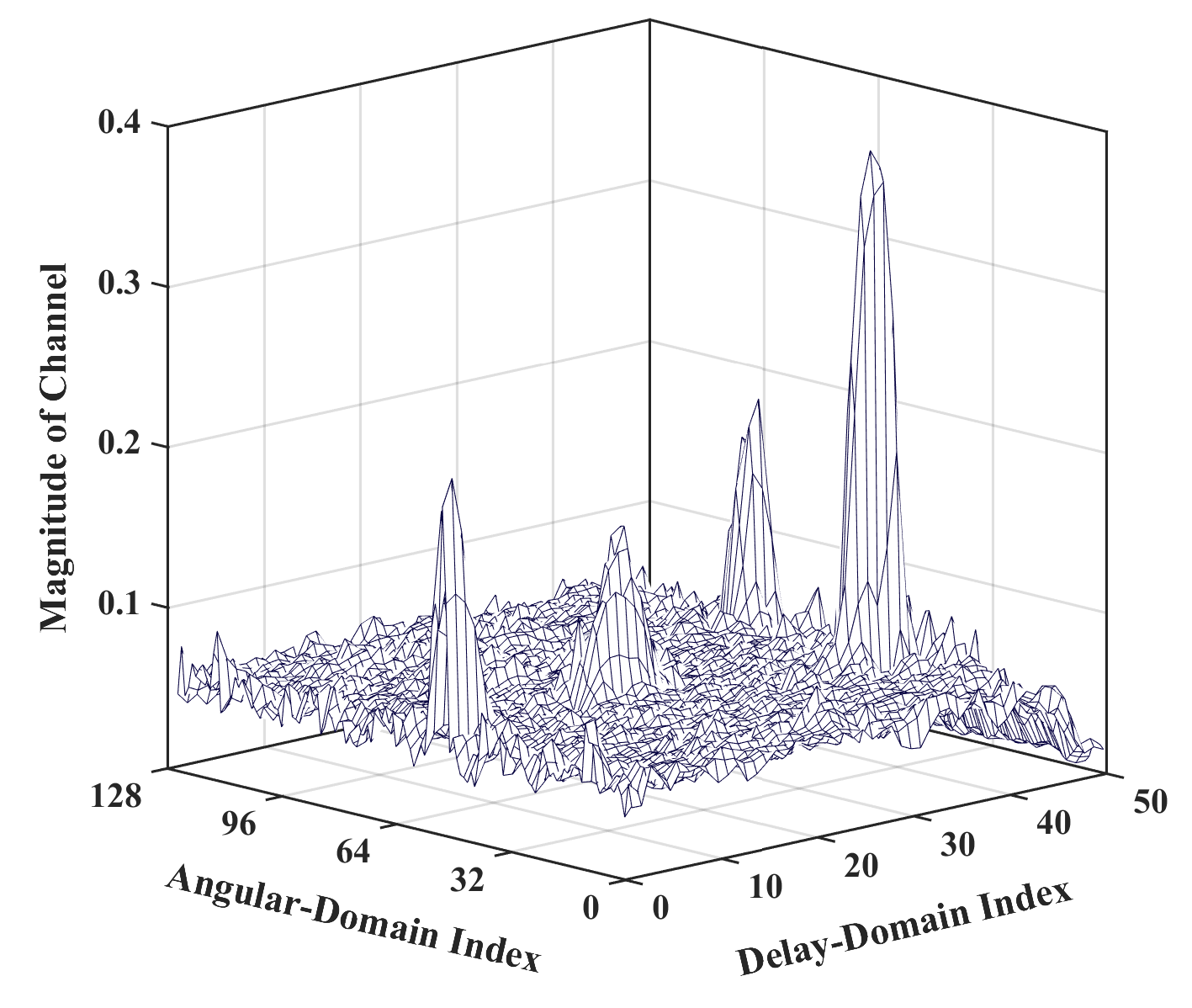}
	\caption{Channel Denoised Output with Unknown Variance.}
	\label{fig_chan_denoised}
\end{figure}

\begin{figure}[h]
	\centering
	\includegraphics[width=0.85\linewidth]{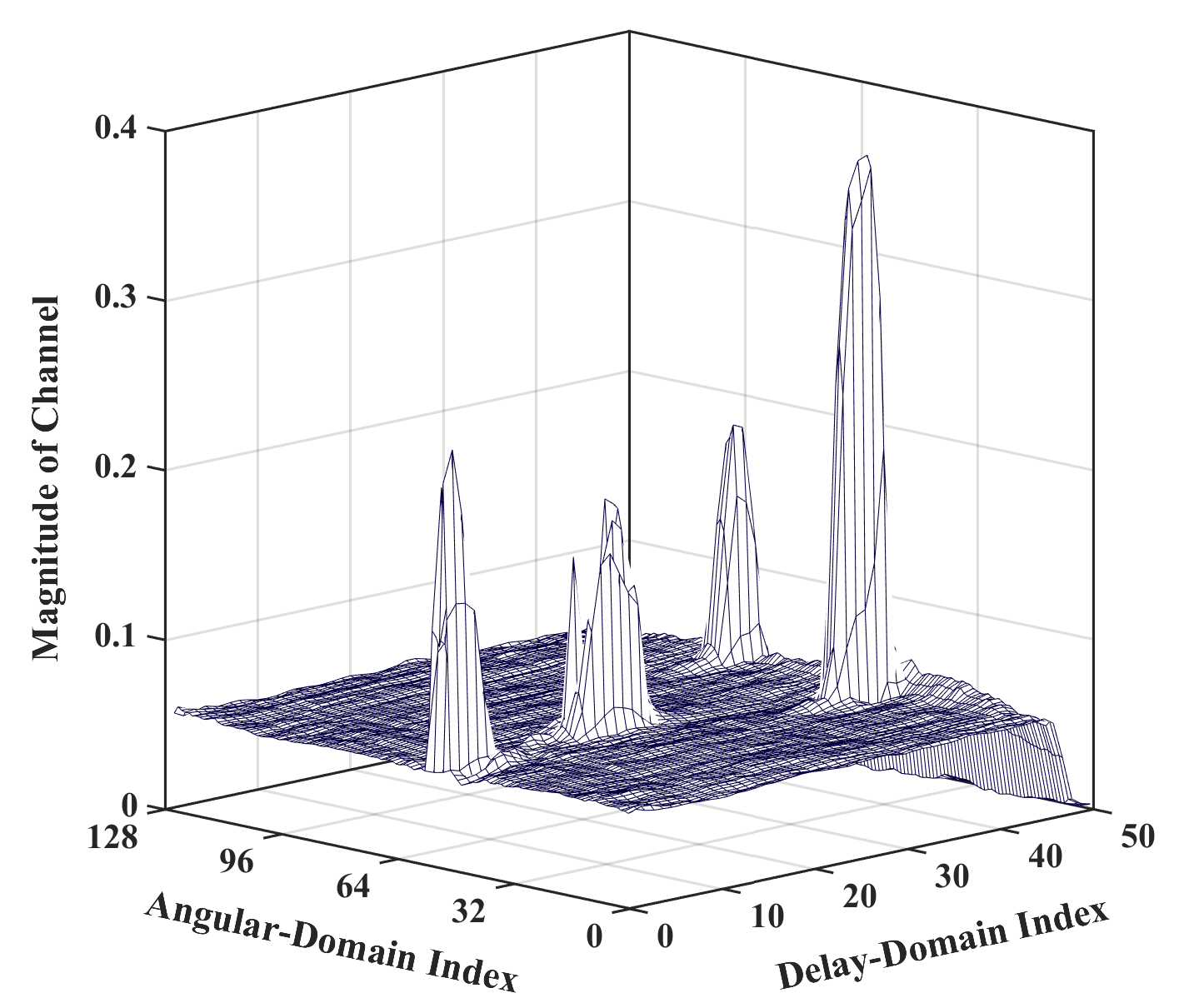}
	\caption{Channel Denoised Output with Known Variance.}
	\label{fig_chan_denoised_known}
\end{figure}

Finally, we re-normalize the denoised image outputs from Figs.~\ref{fig_chan_denoising}(d) and \ref{fig_chan_denoising}(e) back to the CIR domain. The resulting denoised CIRs are shown in Figs.~\ref{fig_chan_denoised} and \ref{fig_chan_denoised_known}, respectively. Compared to the noisy CIR in Fig.~\ref{fig_chan_ls}, the enhanced visibility of the path signatures in Fig.~\ref{fig_chan_denoised_known} clearly demonstrates the effectiveness of the DL-based denoising.

We report the signal-to-noise ratio (SNR) before denoising ($\gamma_{\text{BD}}$) and after denoising ($\gamma_{\text{AD}}$) for mean, median, and DnCNN-based denoising methods, averaged over 1000 Monte Carlo simulations, as shown in Table~\ref{Table:Table_denoising}. From the relative SNR gain, it can be observed that DnCNN significantly outperforms both mean and median filtering techniques, even at very low SNR levels such as $-25$\,dB. It is important to note that, at relatively higher input SNRs, the denoising process can introduce performance degradation. In such cases, the denoising methods effectively over-smooth the CIR, suppressing meaningful features and leading to negative SNR gain. Consequently, denoising proves most beneficial in extremely noisy environments (e.g., when SNR $\leq -5$\,dB), where the signal components are otherwise indistinguishable from the noise.

\subsection{Performance of Clustering:}
 
\begin{figure}[h]
	\centering
	\includegraphics[width=0.98\linewidth]{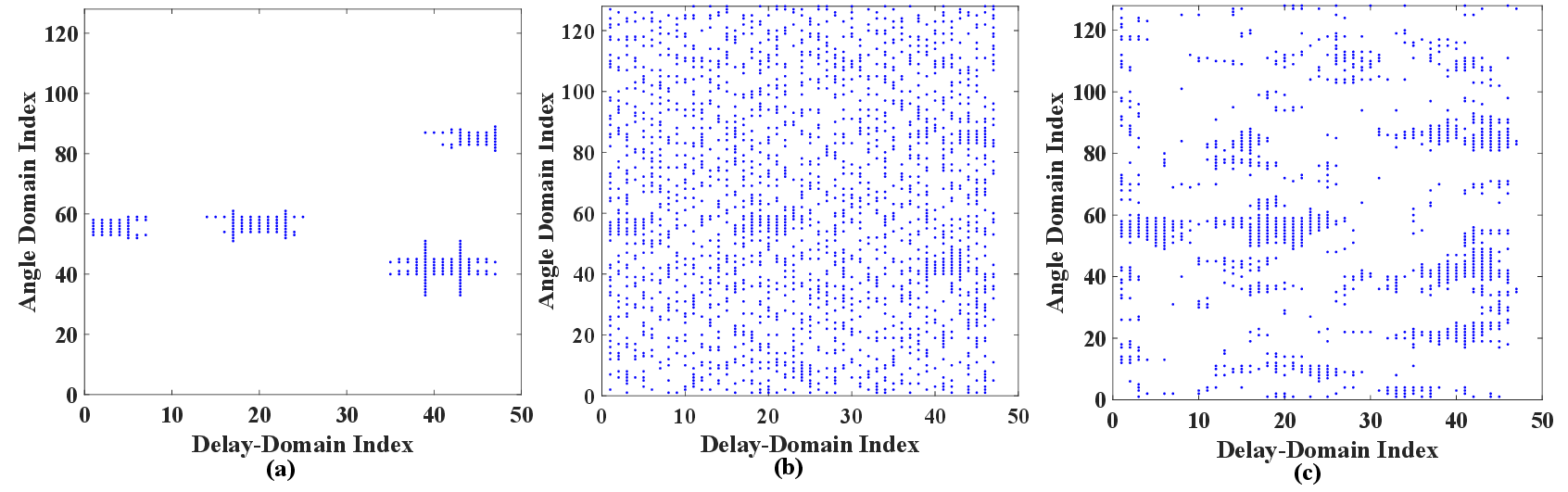}
	\caption{Prepared clustering input using ET (a) without noise (b) -15 dB SNR without denoising (c) -15dB SNR with denoising.}
	\label{fig_chan_clust_et}
\end{figure}

\begin{figure}[h]
	\centering
	\includegraphics[width=0.98\linewidth]{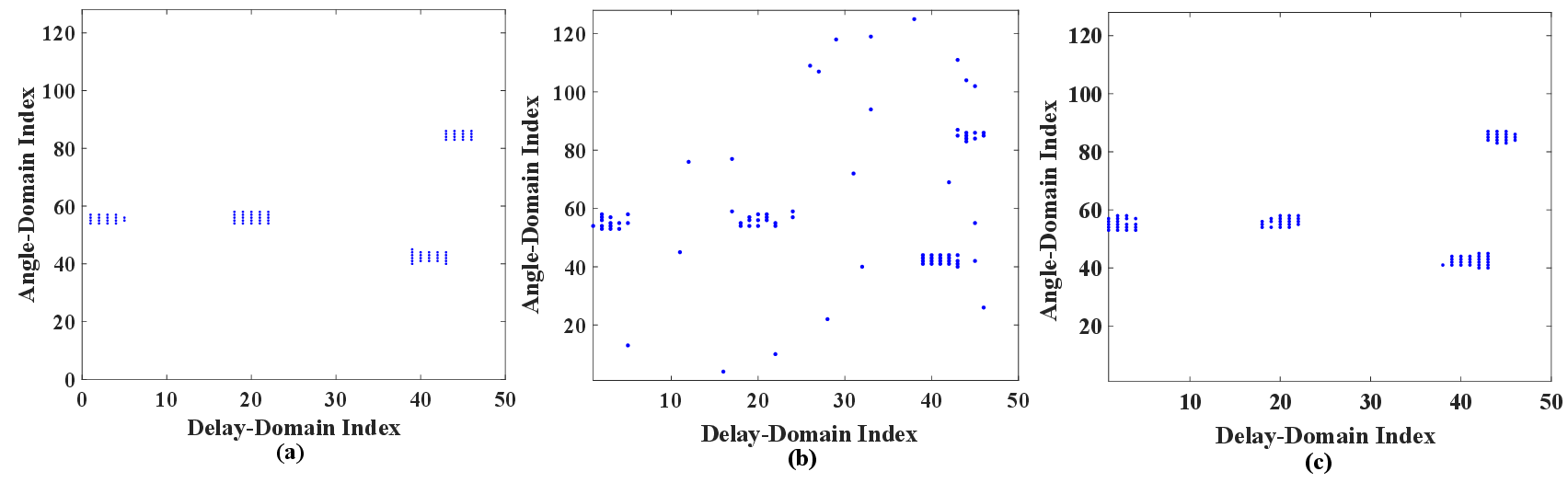}
	\caption{Prepared clustering input using PT (a) without noise (b) -15dB SNR without denoising (c) -15dB SNR with denoising.}
	\label{fig_chan_clust_pt}
\end{figure}

 The denoised CIRs shown in Figs.~\ref{fig_chan_denoised} and \ref{fig_chan_denoised_known} still exhibit residual noise and scattered artifacts, resulting in a non-smooth surface. Therefore, further preprocessing is required to enhance cluster separability. To this end, we apply hard thresholding using either the energy-based or percentile-based criteria, as defined in \eqref{eq:energy_threshold} and \eqref{eq:percentile_threshold}, respectively. For reference, Fig.~\ref{fig_chan_clust_et}(a) shows the clustered dataset derived from a noise-free input CIR using the energy threshold. In contrast, Figs.~\ref{fig_chan_clust_et}(b) and \ref{fig_chan_clust_et}(c) depict the thresholded datasets at an SNR of $-15$\,dB without and with denoising, respectively. It is evident that the dataset in Fig.~\ref{fig_chan_clust_et}(b) fails to reveal any discernible cluster patterns corresponding to the original input signature. On the other hand, Fig.~\ref{fig_chan_clust_et}(c) reveals weak but noticeable traces of the signature structure, confirming the effectiveness of denoising prior to thresholding. Similarly, Fig.~\ref{fig_chan_clust_pt} illustrates the datasets prepared using the percentile threshold. From this analysis, we conclude that energy-based thresholding performs poorly in preparing the dataset for clustering under extremely low SNR conditions. Denoising followed by a well-chosen threshold is essential for reliable cluster identification.

\begin{figure}[h]
	\centering
	\includegraphics[width=0.98\linewidth]{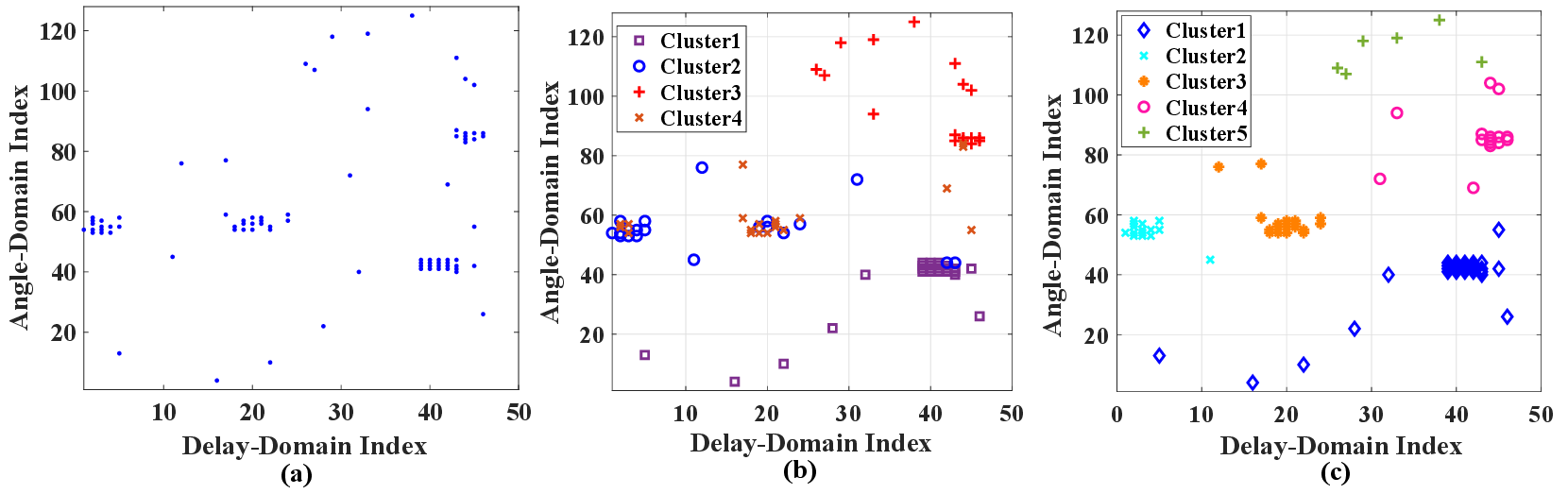}
	\caption{Clustering without denoising (a) Cluster data input using PT (b) K-Means clustering output (c) LGC clustering output.}
	\label{fig_chan_clust_wod}
\end{figure}

  Figure~\ref{fig_chan_clust_wod}(a) shows the result of applying the percentile threshold to the CIR without denoising. Several noise points are retained alongside valid path cluster points. The corresponding clustering results using K-means and the Local Graph Clustering (LGC) algorithm are shown in Figs.~\ref{fig_chan_clust_wod}(b) and \ref{fig_chan_clust_wod}(c), respectively. Due to its assumption of spherical clusters, K-means performs poorly in this setting, producing overlapping clusters and a large spread. In contrast, LGC better captures the structure of the data. The Clustering Metric (CM), which quantifies the compactness of the identified clusters, is $2.5933 \times 10^6$ for K-means and only $4.211 \times 10^4$ for LGC, confirming LGC’s superior robustness in noisy conditions.

\begin{figure}[h]
	\centering
	\includegraphics[width=0.98\linewidth]{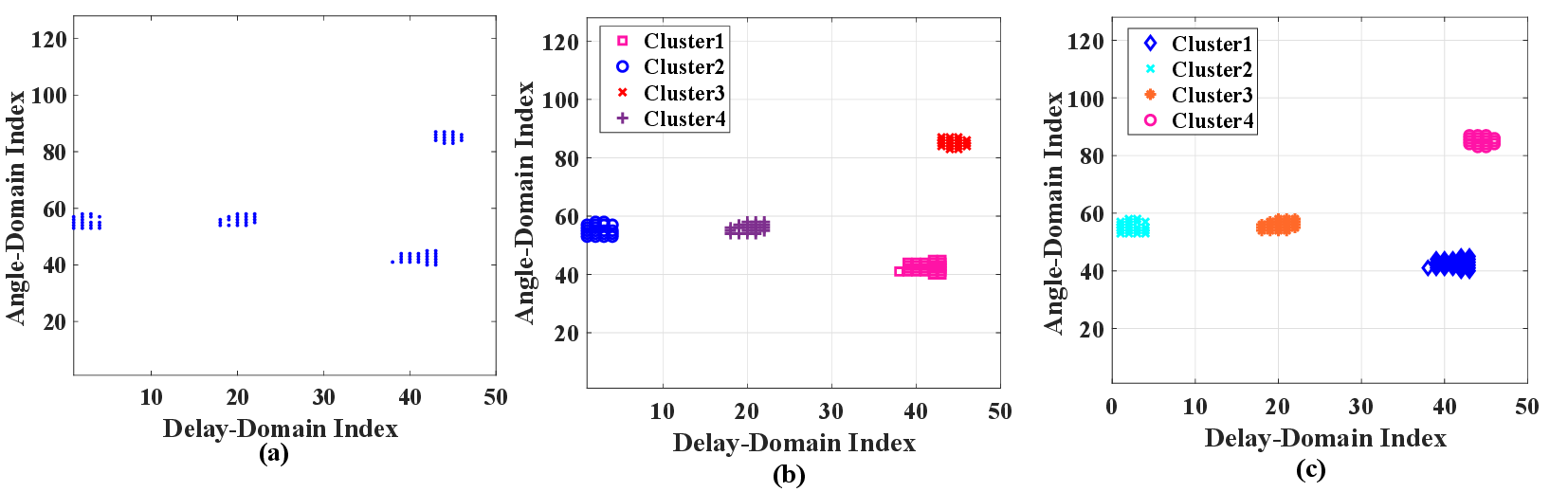}
	\caption{Clustering with denoising (a) Cluster input using PT (b) K-Means clustering output (c) LGC clustering output.}
	\label{fig_chan_clust}
\end{figure}

In Fig.~\ref{fig_chan_clust}(a), the output after applying the percentile threshold to the denoised CIR is shown. The dataset appears clean, with noise effectively suppressed. This step meets our primary objective: preparing a dataset suitable for reliable clustering. As seen in Figs.~\ref{fig_chan_clust}(b) and \ref{fig_chan_clust}(c), both K-means and LGC yield nearly identical clustering results under this clean input. The CM for both methods is low and equal at $14.2512$.

To generalize these observations, we conduct 1000 simulation runs across a range of low SNR values (from 0\,dB to $-25$\,dB) and evaluate clustering quality using the metrics developed in Section~IV. Table~\ref{Table:Table Clustering K-means and LGC} reports the CM values for K-means and LGC under both energy and percentile thresholding. Results indicate that K-means performs poorly with energy thresholding, producing very high CM values, while the percentile threshold becomes effective only below $-15$\,dB. For LGC, energy thresholding results in high CM below 0\,dB, whereas percentile thresholding maintains a low CM even down to $-20$\,dB, demonstrating its robustness in ultra-low SNR scenarios.


 Table~\ref{Table:Table Clustering AE K-Means and LGC} presents the Absolute Error (AE) in the number of clusters identified using K-means and LGC. For K-means, AE is significantly high (approximately 7) when using energy thresholding (ET), but reduces to around 1 when using percentile thresholding (PT) at SNR levels below $-15$\,dB. In contrast, LGC yields a low AE with ET only for SNR values above $-10$\,dB, while PT maintains a low AE even down to $-25$\,dB. To estimate the number of clusters for K-means, we employ the elbow method, which requires specifying a maximum allowable number of clusters ($K_{\text{max}}$). Since K-means forcibly assigns every data point to one of the $K_{\text{max}}$ clusters, improper denoising and thresholding of a noisy CIR can result in overestimation, producing a high AE. LGC, on the other hand, uses a gravitation-based approach that computes local resultant force vectors and centrality-coordination metrics based on a user-defined number of neighbors. This enables LGC to simultaneously determine both the number and structure of clusters. The number of neighbors must be selected judiciously, but overall, LGC tends to produce more accurate estimates of cluster count, especially in noisy conditions.

\begin{figure}[h]
	\centering
	\includegraphics[width=0.98\linewidth]{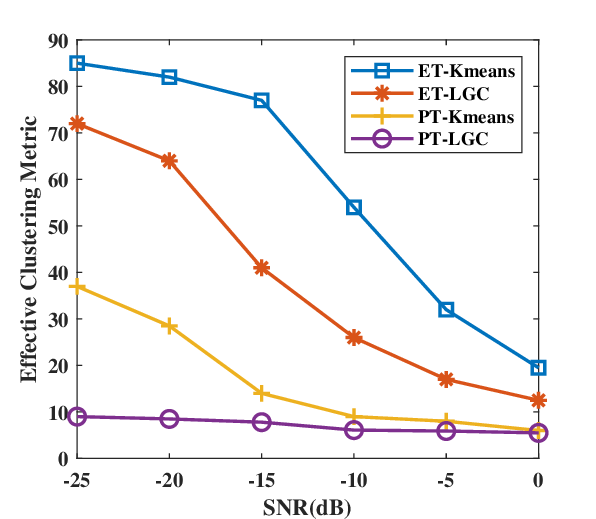}
	\caption{Effective clustering metric for k-Means and LGC with energy threshold and percentile threshold.}
	\label{fig_clust_ae_cm_1}
\end{figure}

As discussed in Section~IV, we use the Effective Clustering Metric (ECM) to evaluate the overall performance of different clustering configurations. The ECM for both K-means and LGC, using both energy and percentile thresholds, is shown in Fig.~\ref{fig_clust_ae_cm_1}. The results confirm that K-means performs poorly with ET across all SNR levels. Interestingly, K-means with PT outperforms LGC with ET, highlighting the advantage of percentile-based thresholding. However, LGC combined with PT consistently provides the best performance across the full SNR range from $-25$\,dB to 0\,dB, making it the most robust and reliable option for both cluster identification and coarse signature estimation.


 \begin{table}[h]
\caption{Clustering Metric for K-Means and LGC}
\label{Table:Table Clustering K-means and LGC}
\centering
\resizebox{\columnwidth}{!}{
\begin{tabular}{|c|cc|cc|}
\hline
\multirow{2}{*}{\textbf{SNR}} & \multicolumn{2}{c|}{\textbf{K-Means}}     & \multicolumn{2}{c|}{\textbf{LGC}}      \\ \cline{2-5} 
                          & \multicolumn{1}{c|}{\textbf{Energy }} & \textbf{Percentile} & \multicolumn{1}{c|}{\textbf{Energy }} & \textbf{Percentile} \\ \hline
                          -25 & \multicolumn{1}{c|}{$ 1.8192 \times 10^{11} $}   &  $3.2181 \times 10^{8}$  & \multicolumn{1}{c|}{$8.5009 \times 10^{10}$}   &  $4.6316 \times 10^{4}$  \\ \hline
                          -20 & \multicolumn{1}{c|}{$ 1.2486 \times 10^{11} $}   &  $7.7578 \times 10^{7}$  & \multicolumn{1}{c|}{$1.2004 \times 10^{15}$}   &  $1.2962 \times 10^{4}$  \\ \hline
                          -15 & \multicolumn{1}{c|}{$ 8.2368 \times 10^{10} $}   & $1.0352 \times 10^{6}$  & \multicolumn{1}{c|}{$2.2378 \times 10^{9}$}   &  $1.4931 \times 10^{3}$  \\ \hline
                          -10 & \multicolumn{1}{c|}{$ 2.3660 \times 10^{10} $}   & $6.1048 \times 10^{4}$  & \multicolumn{1}{c|}{$3.0456 \times 10^{9}$}   &  $220.6407$  \\ \hline
                          -5 & \multicolumn{1}{c|}{$3.8270 \times 10^{9}$}   & $5.6490 \times 10^{3}$  & \multicolumn{1}{c|}{$1.3643 \times 10^{8}$}   &  $140.1065$  \\ \hline
                          0 & \multicolumn{1}{c|}{$3.7515 \times 10^{8}$}   & $334.0516$  & \multicolumn{1}{c|}{$3.3500 \times 10^{6}$}   &  $78.0437$  \\ \hline
\end{tabular}
}
\end{table}

\begin{table}[h]
\centering
\caption{Absolute Error for K-Means and LGC}
\label{Table:Table Clustering AE K-Means and LGC}
\begin{tabular}{|c|m{1.25cm}|m{1.25cm}|m{1.25cm}|m{1.25cm}|} 
\hline
\multirow{2}{*}{\textbf{SNR (dB)}} & \multicolumn{2}{c|}{\textbf{K-Means}}           & \multicolumn{2}{c|}{\textbf{LGC}}               \\ 
\cline{2-5}
                                   & \textbf{Energy } & \textbf{Percentile } & \textbf{Energy } & \textbf{Percentile}  \\ 
\hline
-25                                & $6.5740$                 & $3.3950$                      & $ 2.6770 $              & $0.8130$                       \\ 
\hline
-20                                & $6.4760$                 & $2.6420$                      & $3.2560$                  & $0.8290$                       \\ 
\hline
-15                                & $6.0050$                & $1.3190$                      & $3.3650$                  & $1.0040$                       \\ 
\hline
-10                                & $4.1490$                  & $0.9070$                      & $1.7140$                  & $0.9310$                       \\ 
\hline
-5                                 & $ 2.2820 $              & $0.9520$                      & $0.9740 $               & $0.9320$                       \\ 
\hline
0                                  & $1.2950$                  & $0.8370$                      & $0.8480$                  & $0.8870$                       \\
\hline
\end{tabular}
\end{table}

\subsection{Performance of Fine-Tuning}
\begin{figure}[h]
	\centering
	\includegraphics[width=0.98\linewidth]{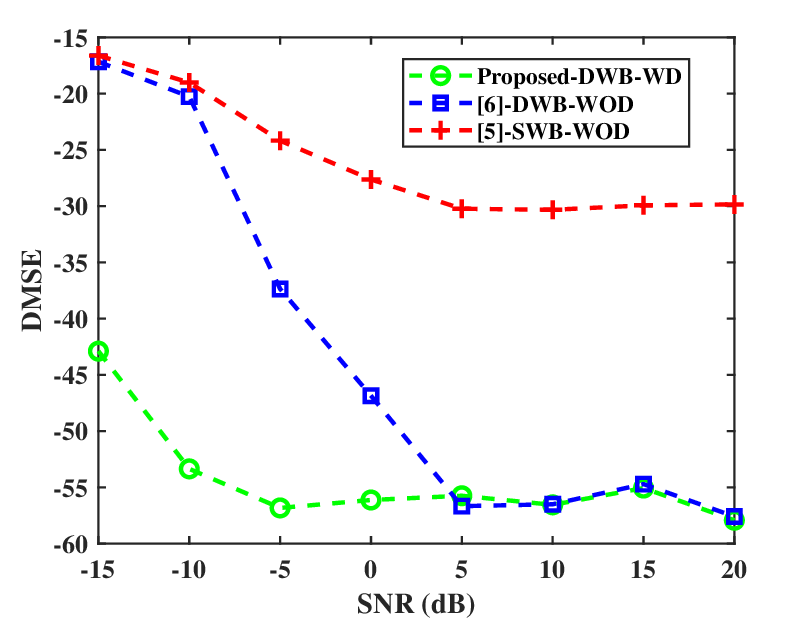}
	\caption{ DMSE of the AI-enabled framework and comparison with the dual wideband without denoising and single wideband without denoising, at M=128, N=128, and $f_s = 0.1f_c$.}
	\label{fig_dev_met_1}
\end{figure}

The effectiveness of fine-tuning is evaluated using the Denoised Mean Square Error (DMSE) metric, as defined in \eqref{eq_DMSE}, with the rotation levels set to $R_M = 15$ and $R_N = 15$. The advantage of our AI-enabled framework is evident in Fig.~\ref{fig_dev_met_1}, where the proposed method achieves significantly lower DMSE compared to the baseline method without denoising \cite{wang2018spatial}, particularly in the low SNR range from $-15$\,dB to $5$\,dB. Additionally, the model in \cite{xie2016unified}, which considers only the temporal wideband effect while neglecting the spatial wideband component, performs poorly even at high SNR values. This highlights the importance of modeling both dimensions in wideband systems.

\begin{figure}[h]
	\centering
	\includegraphics[width=0.98\linewidth]{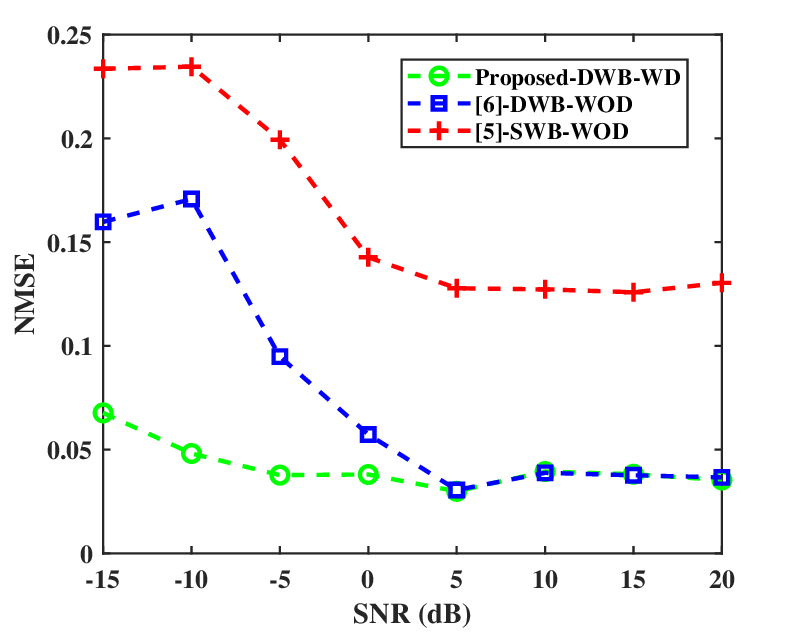}
	\caption{NMSE of the complex path gains $\alpha$ of AI-enabled framework and comparison with the dual wideband without denoising and single wideband without denoising, at M=128, N=128, and $f_s = 0.1f_c$.}
	\label{fig_dev_met_2}
\end{figure}

Figure~\ref{fig_dev_met_2} illustrates the Normalized Mean Square Error (NMSE) of the estimated complex path gains. In the low SNR regime, our method achieves an NMSE improvement of approximately 10\,dB (with NMSE around 0.06), and it asymptotically converges to the performance of \cite{wang2018spatial} for SNR values $\geq 5$\,dB.

\begin{figure}[h]
	\centering
	\includegraphics[width=0.98\linewidth]{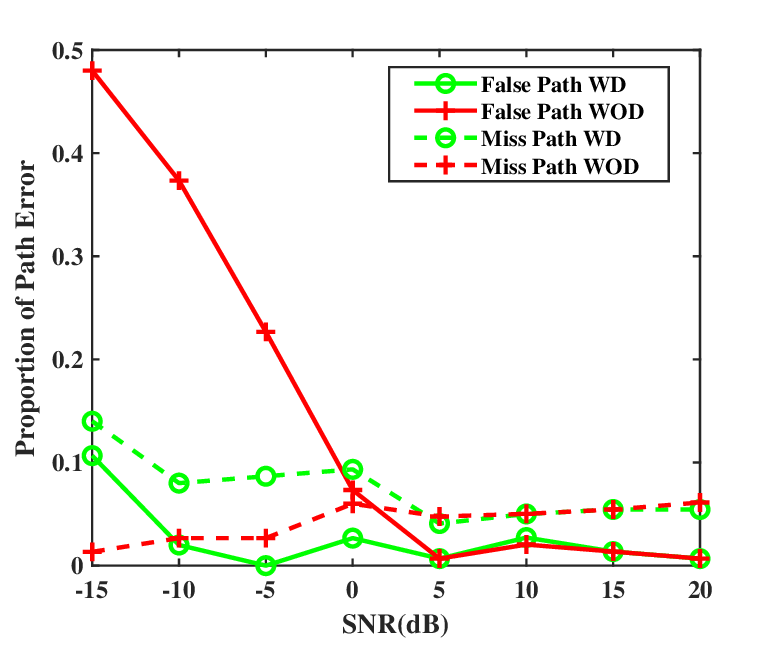}
	\caption{Proportion of paths in error (falsely detected paths and missed paths) with denoising and without denoising.}
	\label{fig_dev_met_3}
\end{figure}

Finally, Fig.~\ref{fig_dev_met_3} presents the proportion of paths that are falsely detected or missed. Without denoising, approximately 48\% of paths are falsely detected; this drops significantly to 10\% with denoising. While denoising introduces a slight increase in the number of missed paths, this trade-off is acceptable. Overall, the proposed framework offers a net gain by successfully recovering many of the true paths that would otherwise remain undetectable in extremely low SNR conditions.
 
\vspace{0.5cm}
\subsection{Complexity Analysis}
We adopt the pre-trained DnCNN configuration as specified in \cite{7839189}. Leveraging GPU acceleration, the proposed framework achieves near real-time performance. Specifically, denoising a $256 \times 256$ delay-angle CIR on an NVIDIA Tesla V100-PCIE GPU takes approximately 0.9067 milliseconds per channel. The computational complexity of the fine-tuning procedure is given by $O(MN \log MN + MN + R_M R_N MN)$, where $R_M$ and $R_N$ denote the resolution levels used during the 2D rotation refinement.

\section{Conclusion}

We have developed a robust signature estimation framework for dual wideband mmWave Massive MIMO OFDM systems operating in extremely high noise environments. Our approach demonstrates the effectiveness of the DnCNN image denoiser in recovering low channel amplitudes even under ultra-low SNR conditions. Leveraging advanced GPU computing, we have optimized the denoiser to function in real-time, achieving a runtime complexity of approximately 1 ms. Additionally, our novel, prior-free gravitation-based clustering method has proven to accurately identify the number of clusters and their respective spreads, outperforming the conventional K-means algorithm. In sparse scenarios, we have shown that percentile thresholding outperforms energy thresholding in preparing the dataset for effective clustering. The end-to-end simulation of our AI-enabled framework reveals exceptional performance in ultra-low SNR regions, surpassing traditional frameworks. Future work could further enhance denoising performance at even lower SNR levels by training deep networks on more realistic channel datasets. Moreover, exploring overlapping clusters for coarse estimation represents an exciting avenue for future investigation.

\ifCLASSOPTIONcaptionsoff
  \newpage
\fi

\bibliographystyle{IEEEtran}
\bibliography{Biblli1}
\end{document}